\documentclass[12pt,preprint]{aastex}

\newcommand{\FUSE}{{\it FUSE}}

\begin{document}
\title{The Abundance of Interstellar Fluorine and Its Implications}

\author{Theodore P. Snow, Joshua D. Destree, and Adam G. Jensen}

\affil{Center for Astrophysics and Space Astronomy}
\affil{Department of Astrophysical and Planetary Sciences, University of Colorado at Boulder}
\affil{Campus Box 389}
\affil{Boulder, CO 80309-0389}

\email{tsnow@casa.colorado.edu, destree@casa.colorado.edu, Adam.Jensen@colorado.edu}

\begin{abstract}
We report results from a survey of neutral fluorine (F I) in the 
interstellar medium.  Data from the {\it Far Ultraviolet 
Spectroscopic Explorer} ({\it FUSE}) were used to analyze 26 lines of 
sight lying both in the galactic disk and halo, including lines to Wolf-Rayet stars and through known supernova remnants.  The equivalent 
widths of fluorine resonance lines at 951.871 \AA\ and 954.827 \AA\ were 
measured or assigned upper limits and combined with a nitrogen 
curve of growth to obtain F I column densities.  These column 
densities were then used to calculate fluorine depletions.  
Comparisons are made to the previous study of F I by 
Federman et al. (2005) and implications for F I formation and depletion are discussed. 
\end{abstract}

\section{Introduction: Fluorine in the interstellar medium}
Fluorine (element 9, with only one stable isotope [$^9$$_{19}$F]) is the most reactive 
species observed in the diffuse interstellar medium.  The diatomic molecule HF is predicted 
to be abundant, as it is formed by an exothermic neutral-neutral reaction of atomic fluorine 
(F I) with H$_2$.  The abundance of HF should compete with, or even dominate over, the abundance 
of F I in molecular clouds, despite the slow reaction rate of F I with H$_2$ (Zhu et al. 2002).  In such 
dense clouds, F I also can be converted rapidly into CF$^+$ (Neufeld, Wolfire, \& Schilke 2005). 
But in diffuse molecular clouds (i.e., regions where some of the hydrogen, but not all, is in 
molecular form; see Snow \& McCall 2006), HF should still be detectable but not dominant over 
F I. In these clouds, F I is observed in the gas phase, as demonstrated previously by Snow and 
York (1981) and Federmen et al. (2005).

There are three possible formation mechanisms for fluorine: (1) Type 
II supernovae, through a neutrino capture process starting from 
$^{20}$Ne, typically called the $\nu$-process (Woosley \& Haxton, 1988 and Woosley \& Weaver, 1995), (2) Wolf-Rayet stars, 
through internal helium burning followed by rapid ejection in stellar 
winds (Meynet \& Arnould, 2000), and (3) in asymptotic giant branch (AGB) stars, again through helium burning followed by 
mass loss (Forestini et al. 1992).  In all three scenarios rapid mass loss is required, 
as fluorine is destroyed in thermodynamic equilibrium 
almost as soon as it forms, through $\alpha$-captures leading to 
$^{16}$O and $^{22}$Ne. Thus the galactic abundance of fluorine might 
help to distinguish among the possible formation mechanisms, which in 
turn can improve our understanding of the nucleosynthesis history of 
the Sun and other stars.  However, for various reasons, this element 
has not been widely observed in its atomic form despite its implications for the 
history of nucleosynthesis in the Galaxy.

Some observational evidence regarding fluorine formation, especially 
concerning AGB stars, has been reported since the various formation mechanisms were 
proposed.  Though some of the findings seem to conflict, 
most support AGB stars as a fluorine production site.  Jorissen, Smith and 
Lambert (1992) used rotation-vibration lines of HF to obtain fluorine abundances 
in red giants.  They found an overabundance of fluorine in C-rich stars, 
indicating that fluorine could indeed be produced in AGB stars.  Similarly, 
Cunha et al. (2003) used HF rotation-vibration lines to study fluorine in the 
Large Magellanic Cloud and $\omega$ Centauri.  In general, $\omega$ Centauri 
is largely enriched in elements formed through the s-process in AGB stars. This study, however, found low 
F/O ratios in two $\omega$ Centauri stars with low metallicities, indicating that fluorine 
was not enriched along with the s-process elements in the cluster. Thus, it was concluded that 
AGB stars do not play a dominant role in the formation of fluorine.  
Soon after, though, more evidence for fluorine formation in AGB stars was 
reported by Werner, Rauch, and Kruk (2005), as they detected great fluorine 
overabundances using absorption lines from highly ionized fluorine in 
{\it Far Ultraviolet Spectroscopic Explorer} ({\it FUSE}) spectra.  
Recently, Federman et al. (2005) studied fluorine in the ISM towards two stars 
in the Cep OB2 association using the 954 \AA\ F I line in {\it FUSE} data.  
They found no indication of fluorine enhancement resulting from Type II supernovae.  
A study by Zhang and Liu (2005) observed emission lines in a sample 
of planetary nebulae and found fluorine to be generally overabundant, providing 
more evidence for AGB stars playing an important role in fluorine nucleosynthesis.  
Additionally, they saw a large fluorine enhancement in NGC 40, a planetary nebula 
whose central star is a Wolf-Rayet star, suggesting that rapid mass loss as seen 
in Wolf-Rayet stars favor fluorine formation. 

Improved modeling of these formation mechanisms can also help answer the questions behind 
fluorine nucleosynthesis.  A recent model by Renda et al. (2005) takes into account fluorine production 
from all three possible sources.  They find that a model Milky Way including all three 
fluorine production mechanisms most closely matches the current observed fluorine abundances. 

In this study we use {\it FUSE} far-ultraviolet spectra to determine 
the interstellar F I abundances in several lines of sight, with the 
hope of helping to constrain the origin of fluorine in the universe. 
The abundance of fluorine in the diffuse ISM can be 
measured through its pair of ground-state atomic F I absorption 
lines in the far-ultraviolet, at wavelengths of 951.871 \AA\ and 
954.827 \AA\ (though usually only the 954 \AA\ line is detected).  
In dense clouds, some transitions of fluorine-based compounds can be 
observed as well (Neufeld, Wolfire, \& Schilke, 2005).  The HF molecule was observed in the ISM 
using the J = 2-1 rotational line by Neufeld et al. (1997) and CF$^+$ was just recently 
discovered by Neufeld et al. (2006) toward the Orion Bar region through emission from three rotational transitions.

In this paper we present the results of a {\it FUSE}-based survey of 
F I column densities in a sample of 26 moderately reddened 
stars, selected for maximum gas column density observable below the 
threshold where the F I features become totally obscured by molecular 
hydrogen absorption.  Our sample stars have E$_{B-V}$ values ranging 
from 0.17 mag to 0.62 mag and total hydrogen column densities in the 
range from 0.4 to 3.2 X 10$^{21}$ cm$^{-2}$.  This is the 
most extensive study of fluorine in diffuse interstellar clouds yet 
attempted.

The first detection of interstellar F I was reported by Snow and York 
(1981) on the basis of {\it Copernicus} spectra of $\delta$ Scorpii, 
but thereafter no one took up the pursuit until the recent {\it 
FUSE}-based study by Federman et al. (2005).  Federman et 
al. observed the 954 \AA\  line of F I in two stars in the Cepheus 
OB2 association and found evidence of slight ($\sim$45\% ) fluorine depletion 
relative to the abundance of this element in the Sun and in 
meteorites.  Data on both can be found in Anders \& Grevesse, 1989; 
Lodders, 2003; and Asplund, Grevesse, \& Sauval, 2005 --- all of which 
agree on the solar fluorine abundance to within $\pm$0.1 dex. 
Federman et al. provide extensive discussion of the 
nucleosynthetic implications of the interstellar fluorine abundance, 
which will not be repeated here except for comments in our discussion 
section.

In the following we present a description of the {\it FUSE} data and 
the rationale for our choice of targets from the archive (\S 2), a description of our methods for analyzing the data (\S 3), and a discussion of the results and their implications (\S 4).

\section{Observations and Data Reduction}

The {\it FUSE} mission 
continues to be a vital tool for analyzing the processes in 
interstellar clouds.  This is largely due to the fact that several of 
the most abundant elements in space have their ground-state 
transitions in the far ultraviolet, a wavelength region covered 
uniquely by {\it FUSE} but not by other ultraviolet (UV) instruments such as the 
{\it Hubble Space Telescope} ({\it HST}).  {\it FUSE} is also more sensitive in 
throughput and/or spectral resolving power than previous instruments 
operated in the same wavelength interval, such as {\it Copernicus} 
or {\it ORFEUS}.

Stars in this study of fluorine were selected from the \FUSE\ archive primarily 
from programs P101 and P102.  We selected targets based on their 
ratios of signal to noise and on the existence of sufficient flux in 
the region immediately surrounding the 954 \AA\ F I 
line.  Table \ref{tab_stardata} shows the properties of all the 
analyzed lines of sight.

For all lines of sight, raw data were downloaded and processed with 
version 2.4.0 of the CALFUSE pipeline.
All the observations were broken down into multiple exposures; so a 
cross-correlation analysis was performed on all detector 
segments before combining the spectra.

\section{Data Analysis}

In order to convert the observed F I absorption line strengths into 
column densities, the following steps were taken: (1) we removed the influence of molecular hydrogen absorption bands which 
partially overlie the F I lines, then (2) we developed a suitable 
curve of growth to take into account possible saturation of the F I lines.
  The first of these steps required us to model and 
remove the H$_2$ absorption; the second required us to develop a 
rationale for deducing the curve of growth that is most applicable to 
the F I lines.

\subsection{$H_2$ Modeling}

For nearly every target, molecular hydrogen lines of the J = 0 and J 
= 1 rotational states interfere with the region surrounding the 954 
\AA\ F I line.  Thus, in order to determine the continuum and to 
measure the equivalent width of the fluorine line, we had to correct 
for the obstruction from H$_2$.  This was done by creating a model of 
the molecular hydrogen absorption lines for the J = 0, 1, and 2 
rotational states.  For lines of sight where a H$_2$ curve of growth 
analysis had already been performed, we took column densities from the 
literature, primarily Shull et al. (in preparation), for our sample.

For targets that had no available H$_2$ data, column densities were 
derived by fitting the 4-0 (1046-1054 \AA ), 2-0 (1075-1082 \AA ), and 1-0 (1091-1097 \AA )  Lyman bands of the spectrum in the 
LiF1A, LiF2A, and LiF2B segments.  We first divided out any 
absorption features in the region around the Lyman bands that were 
not H$_2$ lines of J = 0, 1, or 2 by fitting any 
obvious absorption features with a single Gaussian curve and then dividing 
the spectrum flux by that curve.  A profile-fitting 
procedure was then used to derive the H$_2$ column densities. For 
details on this procedure see Rachford et al. (2002).  Because a 
curve of growth analysis was not performed for H$_2$, we did not 
derive a {\it b}-value but used a value of $b = 5$ $km s^{-1}$ for 
making the models.  We can make this assumption 
because the J = 0 and 1 lines are usually heavily damped and thus are not 
sensitive to fine velocity structure or {\it b}-value.  The J = 2 
line is affected by our choice in {\it b}-value.  However, since this 
line is not critical in our analysis and is only used for aligning 
the spectra, the value is not important.  Derived H$_2$ column 
densities can be found in Table \ref{tab_H2colden}.

For HD 208440 and HD 209339 our H$_2$ column densities and those from Federman et al. 
(2005) generally agree within the 1-$\sigma$ errors.  The one discrepancy is 
the J = 2 column densities for HD 209339 which agree within 3-$\sigma$. This could be due in 
part to the very small error bars on the Federman et al. measurements.  
Error in H$_2$ column density for our study is the standard deviation of the 
individual fits of the three Lyman bands.  This discrepancy in the J = 2 column density, though, does not 
in any way affect our analysis of F I because, as stated before, the J = 2 line 
is only used for aligning the model with our spectra.

\begin{center}
\begin{figure}
\begin{center}
\scalebox{.5}[.5]{\includegraphics[angle=90]{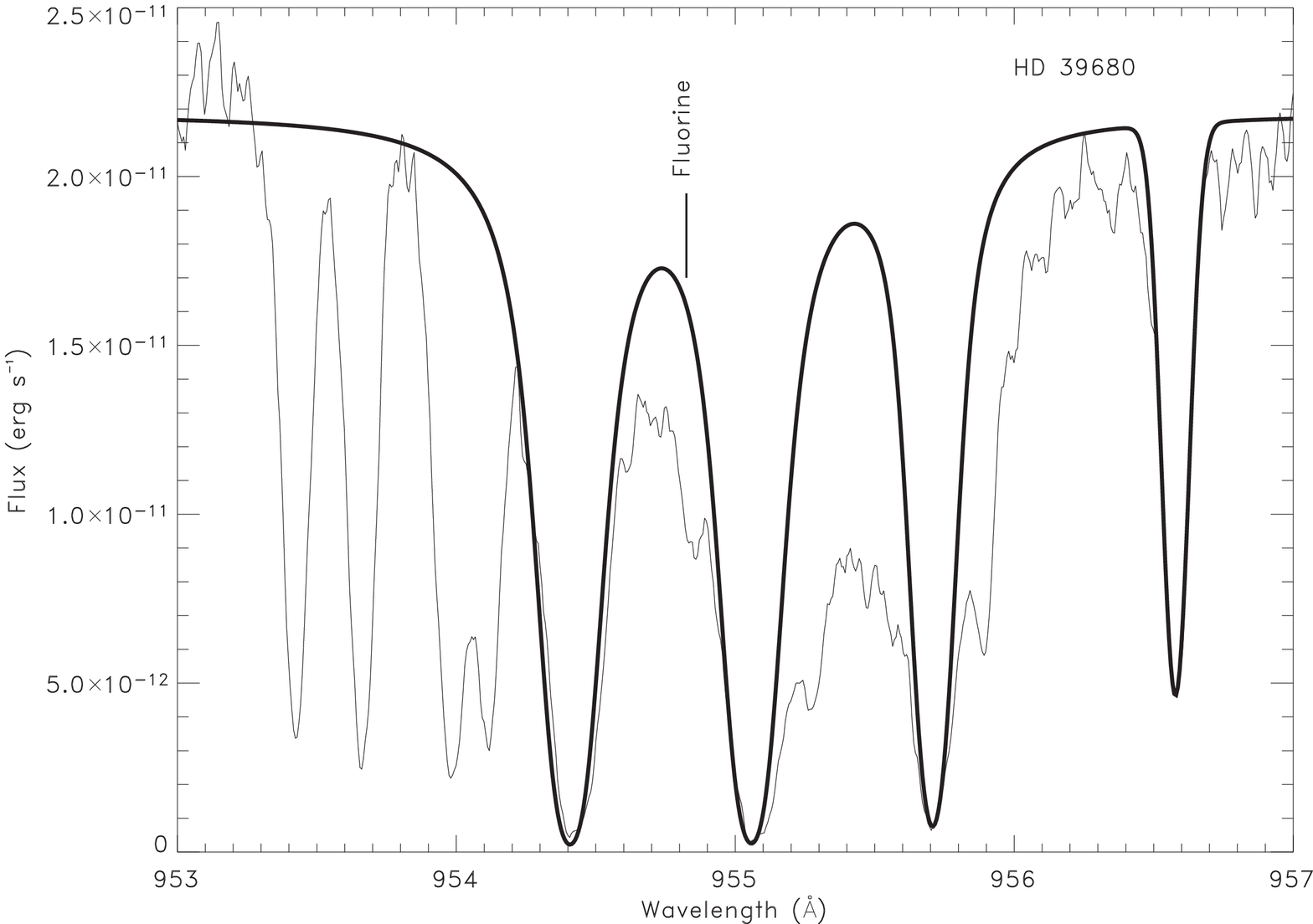}}
\caption{Stellar spectrum for HD 39680 in the region around the 954 \AA\ F I line with the derived $H_2$ model overlaid (bold).\label{fig_modeloverlay}}
\end{center}
\end{figure}
\end{center}

Once we had a model of the molecular hydrogen, we performed a 
cross-correlation analysis on the region around the 
F I line to align the model and spectrum (see Figure \ref{fig_modeloverlay}).  The H$_2$ model was 
correlated with the {\it FUSE} spectrum in the 954 \AA\ to 957 \AA\ 
region, relying heavily on the 956 \AA\ J = 2 hydrogen line. Once 
aligned, the spectrum and corresponding error were divided by the 
hydrogen model to restore the continuum around the F I line.

\subsection{Line Velocities}
In assessing the F I line, whether present or absent, we needed to know where to look for it, which depends on the line of 
sight velocity structure.  This is not a simple question because 
there are noticeable shifts between the atomic nitrogen (N I) and
 H$_2$ absorption lines in many cases.  We expect the distribution 
of atomic fluorine to correlate to that of nitrogen because of the 
similarities in the two elements' ionization potentials.  Thus, we are able to determine the expected F I line 
center based on nearby nitrogen features.

To confirm our assumption that F I and N I should arise at the 
same velocity, we also checked the atomic oxygen (O I) velocity in the two sight lines 
where N I and H$_2$ have very different velocities.  We used O I as confirmation because, like fluorine, the ionization potential of oxygen is close to that of nitrogen.  In both cases we 
found that the O I lines were shifted the same amount from H$_2$ as 
the N I lines were.  Thus, we felt confident that similar shifts for F I exist.

\subsection{F I Equivalent Width Measurements}
\begin{center}
\begin{figure}
\begin{center}
\scalebox{.5}[.5]{\includegraphics[angle=90]{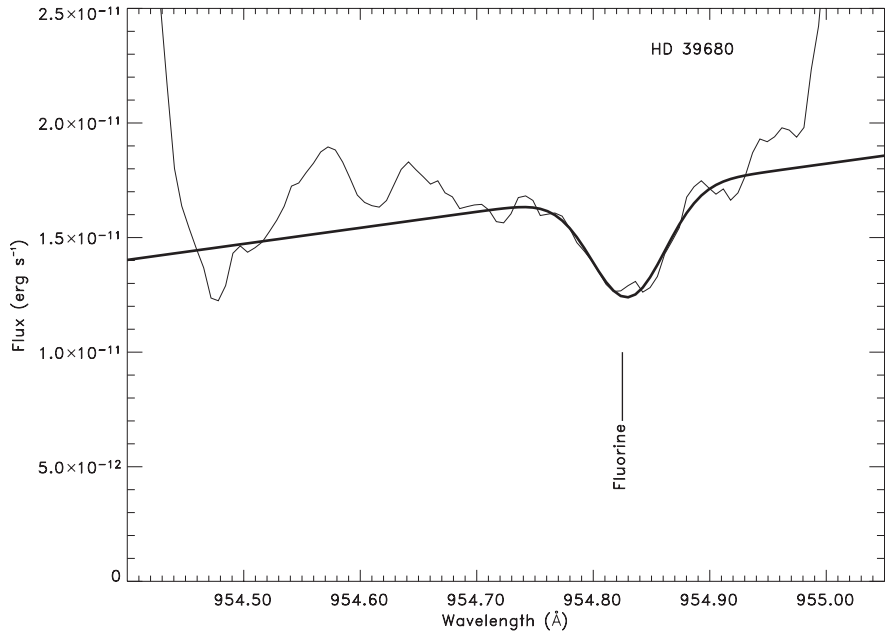}}
\caption{Fitting of 954 \AA\ FI line with a single Gaussian after dividing out the $H_2$ for HD 39680.\label{fig_fitting}}
\end{center}
\end{figure}
\end{center}

Once the spectrum was corrected, we obtained an equivalent width for both
the F I lines by performing simple single Gaussian fits with 
second-order polynomials to fit the nearby continua (see Figure \ref{fig_fitting}).  In cases where the equivalent 
width for a F I lines was not significant to two standard deviations, we 
give a 2$\sigma$ upper limit.  Measured equivalent widths, upper 
limits, and errors are presented in Table \ref{tab_Feqwid}.

\subsection{N I Curve of Growth}

Since we had only one F I line in most cases, we could not derive an 
empirical curve of growth from F I alone.  Instead, based on the 
similarities between the expected distribution of N I and F I in 
space, we have assumed that the curve of growth for F I will be 
very similar to that for N I.

\begin{center}
\begin{figure}
\begin{center}

\scalebox{.5}[.5]{\includegraphics{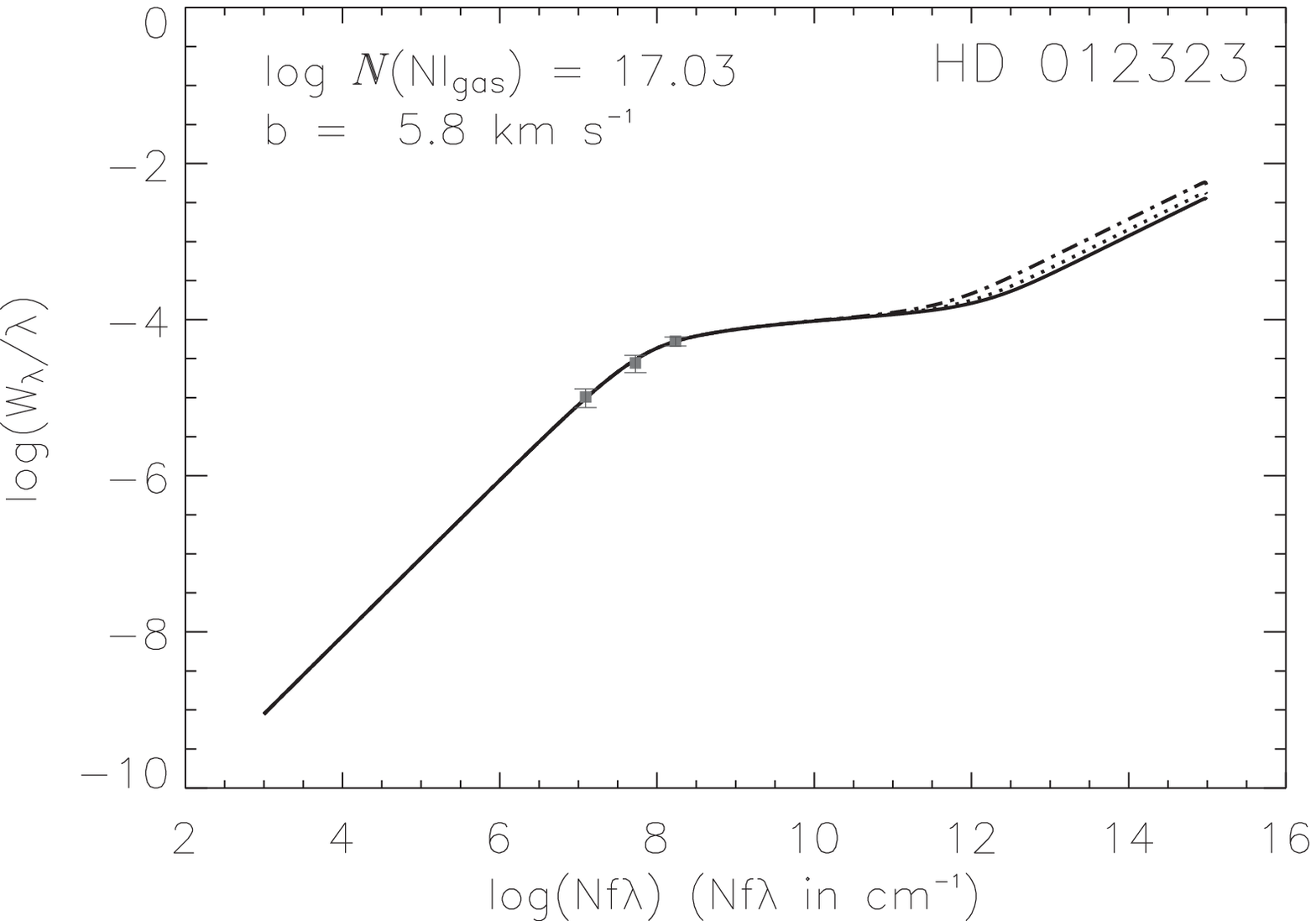} \quad \includegraphics{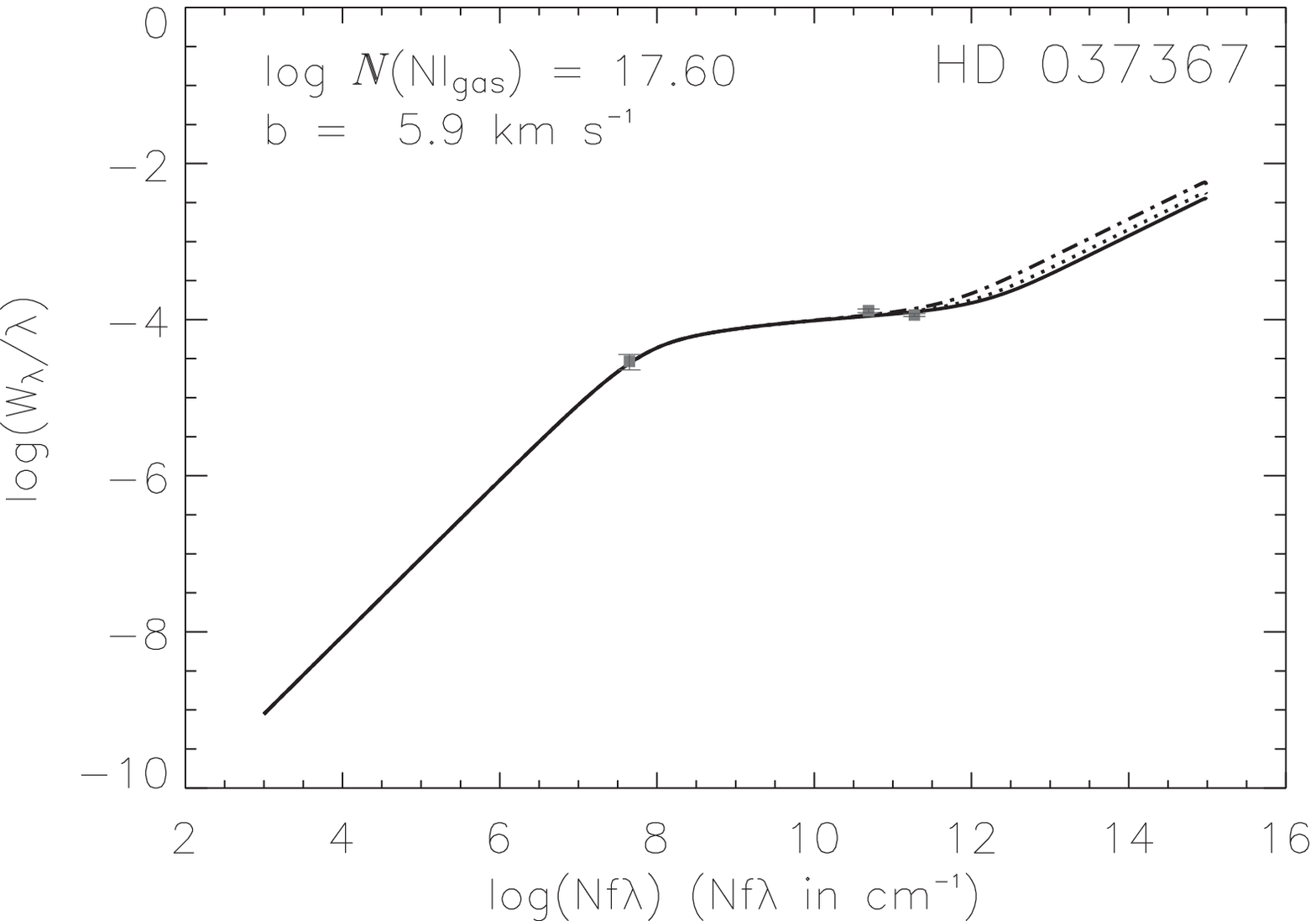}} 

\vspace{.3 cm}

\scalebox{.5}[.5]{\includegraphics{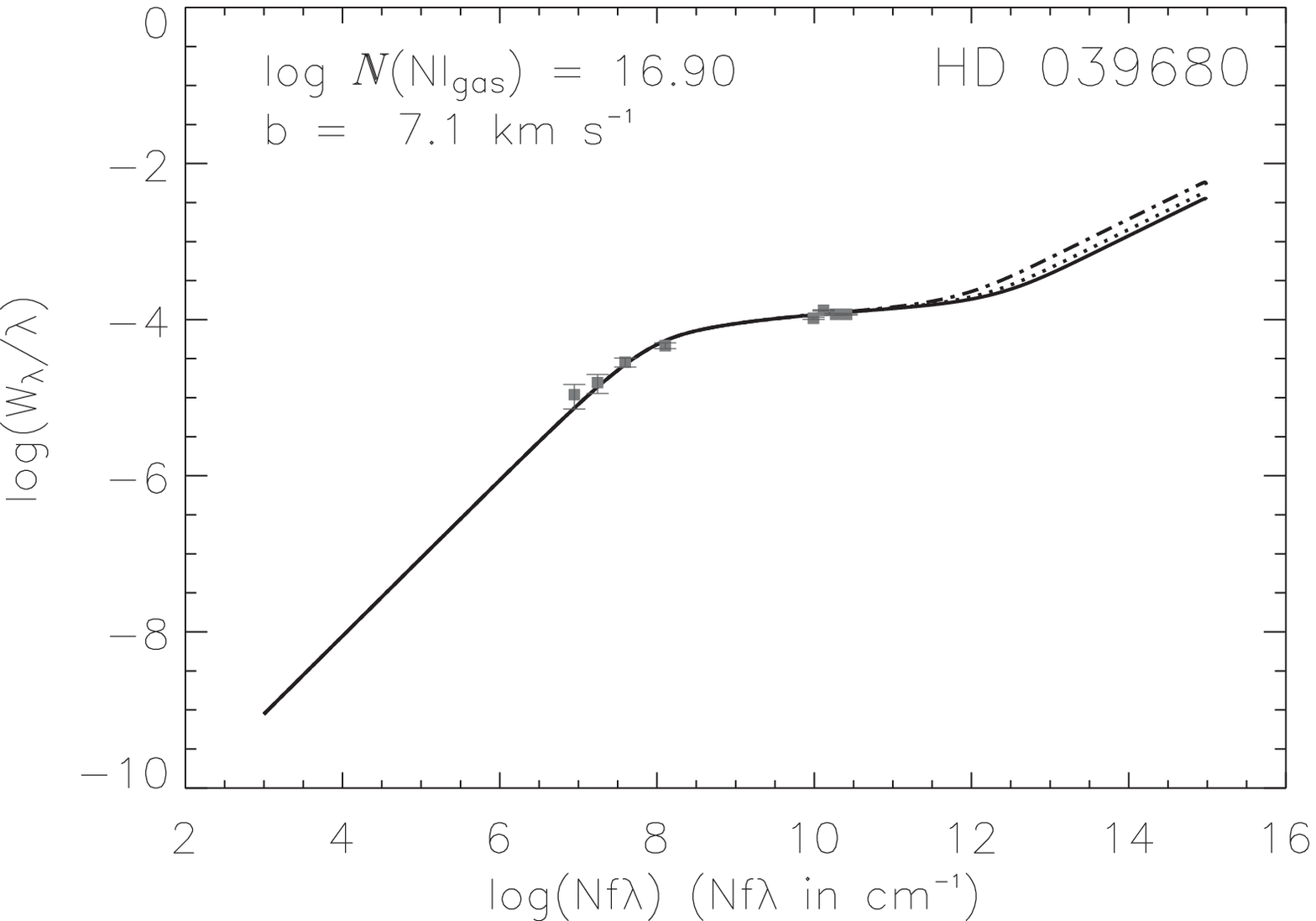} \quad \includegraphics{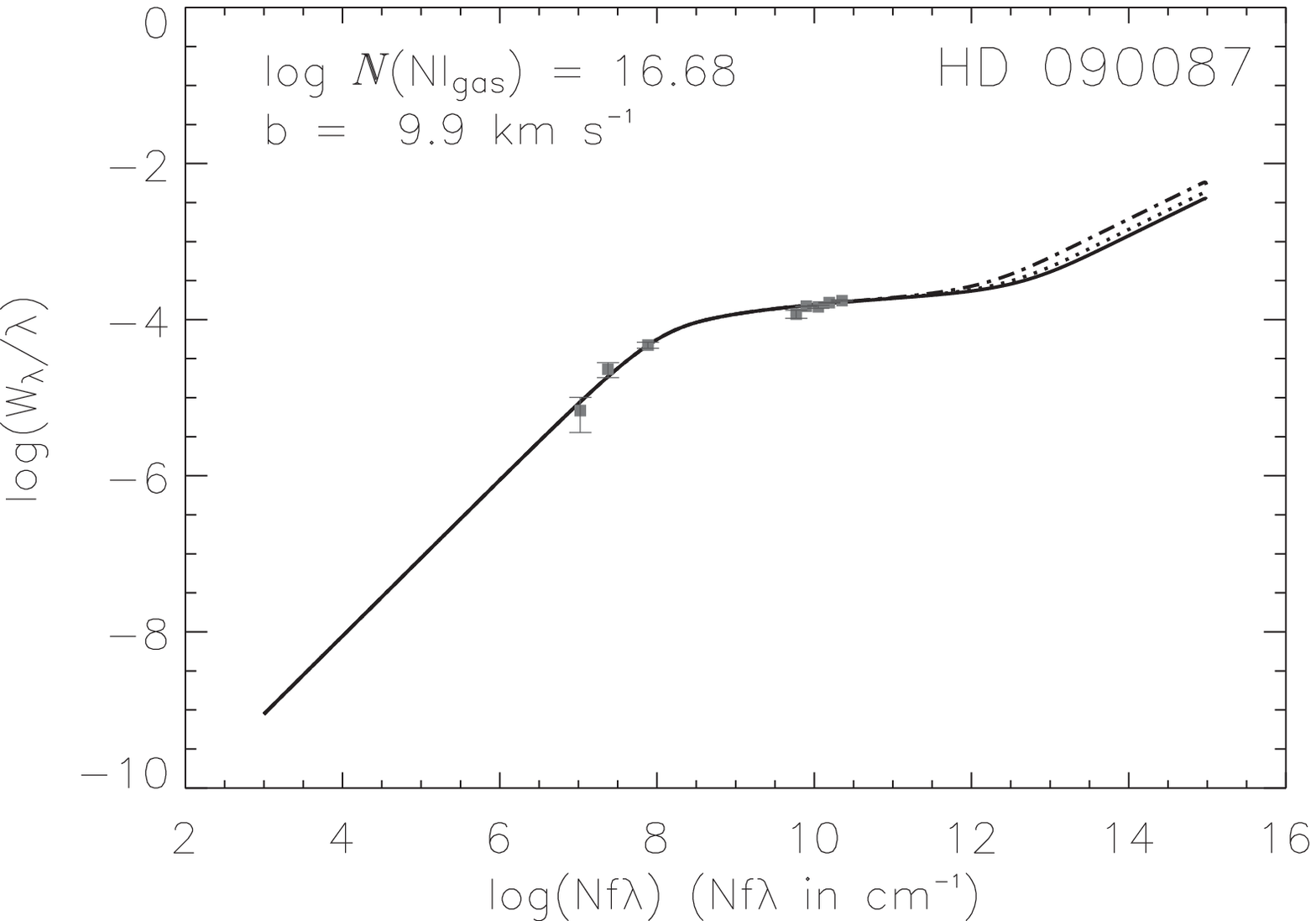}}

\vspace{.3 cm}

\scalebox{.5}[.5]{\includegraphics{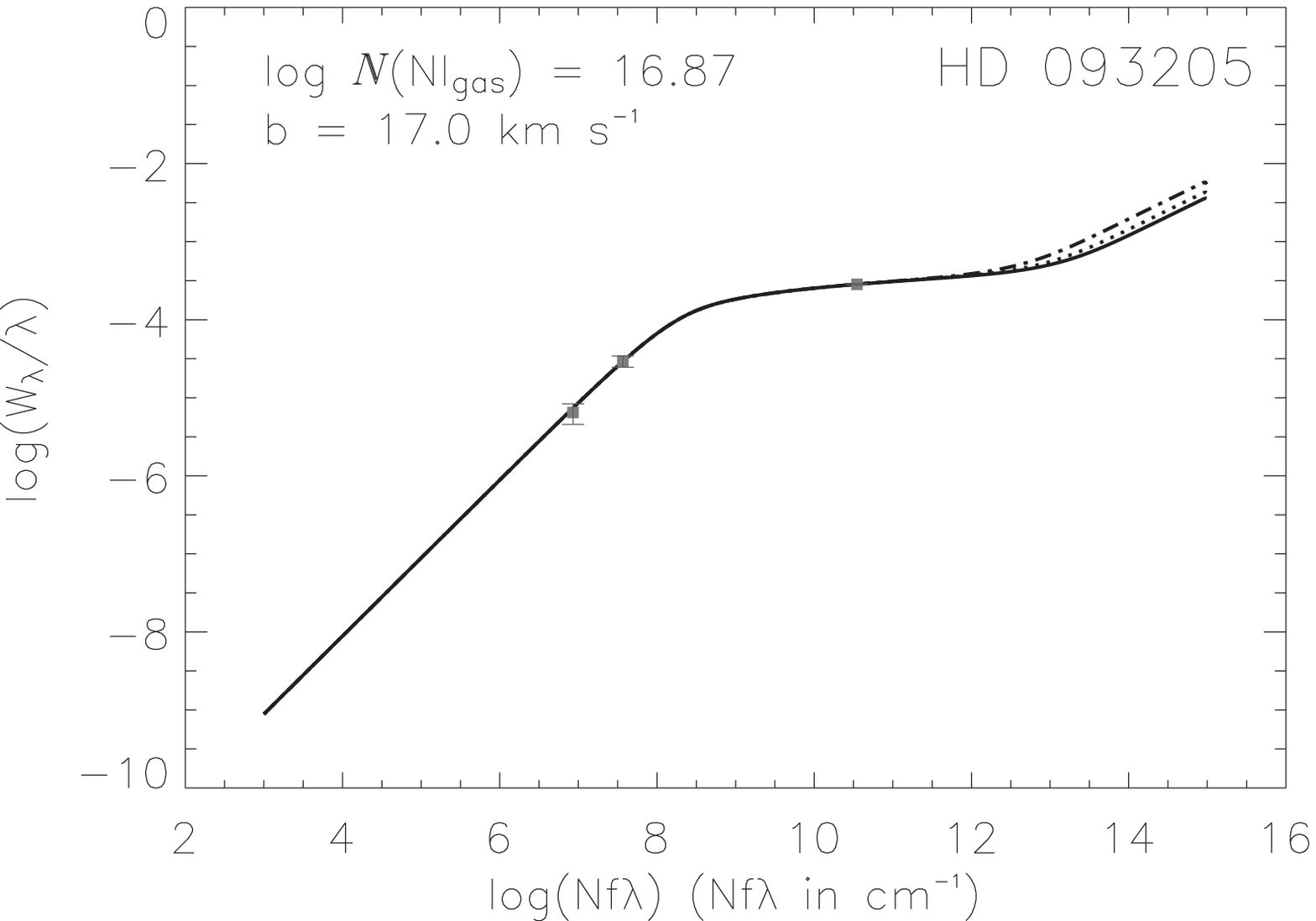} \quad \includegraphics{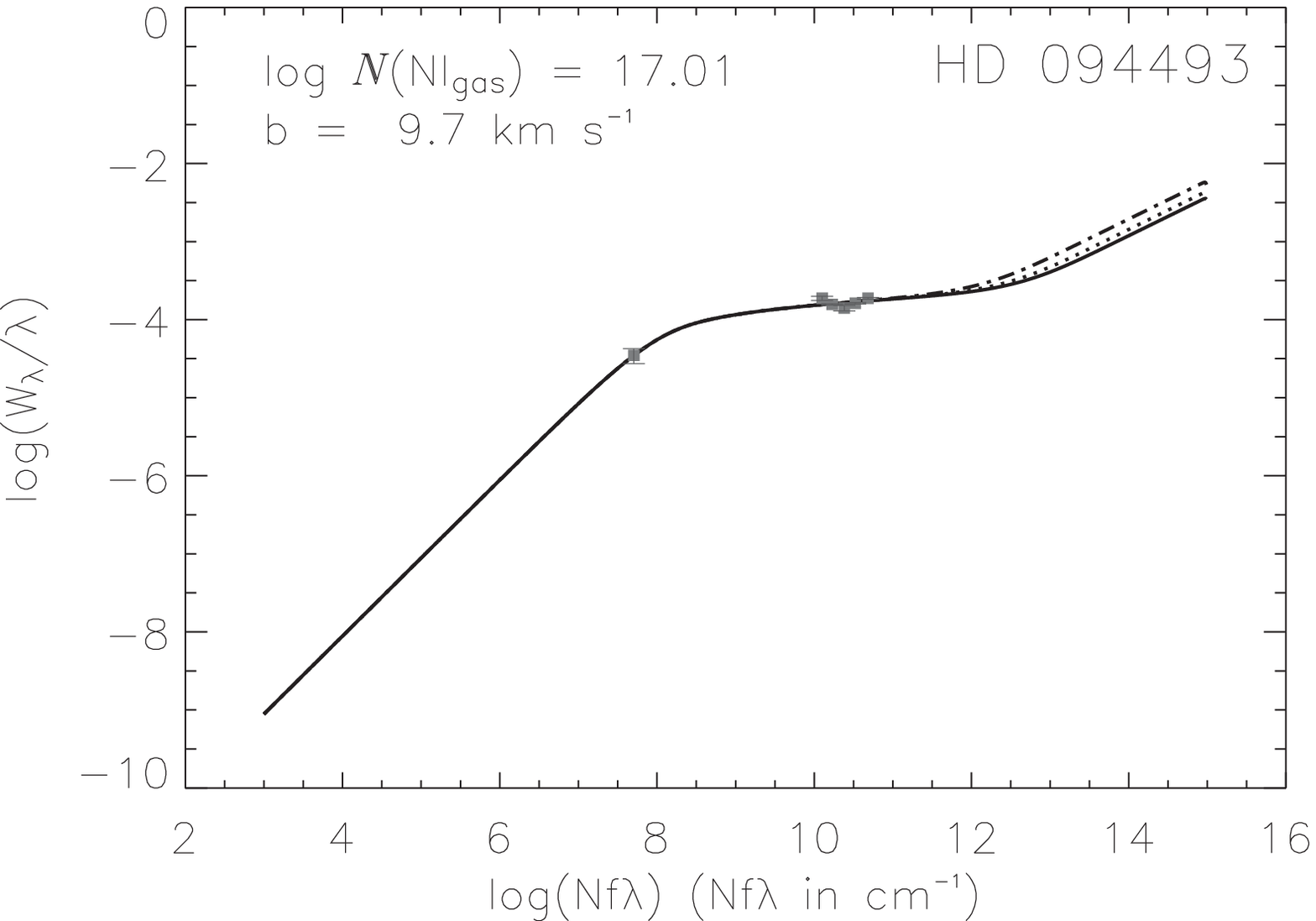}}

\caption{N I curves of growth for HD 12323 through HD 94493.  Measured equivalent widths and error bars are plotted on top of the appropriate curve for the derived column density and {\it b}-value.  The solid, dotted, and dotted-dashed lines in the damping portion of the curve represent damping constants of $1.51\times10^8$, $2.19\times10^8$, and $4.07\times10^8$, respectively.\label{fig_Ncog1}}
\end{center}
\end{figure}
\end{center}

\begin{center}
\begin{figure}
\begin{center}

\scalebox{.5}[.5]{\includegraphics{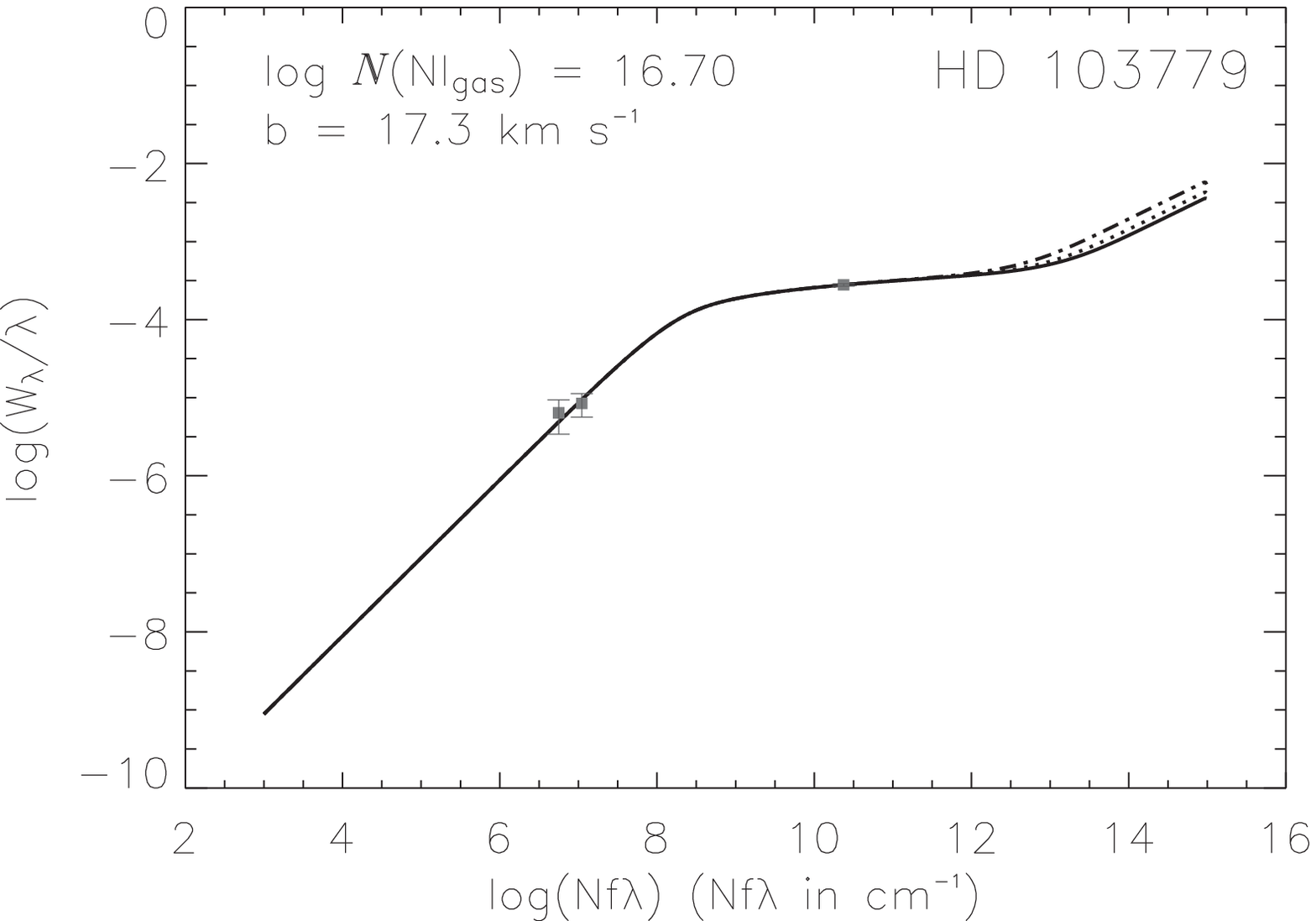} \quad \includegraphics{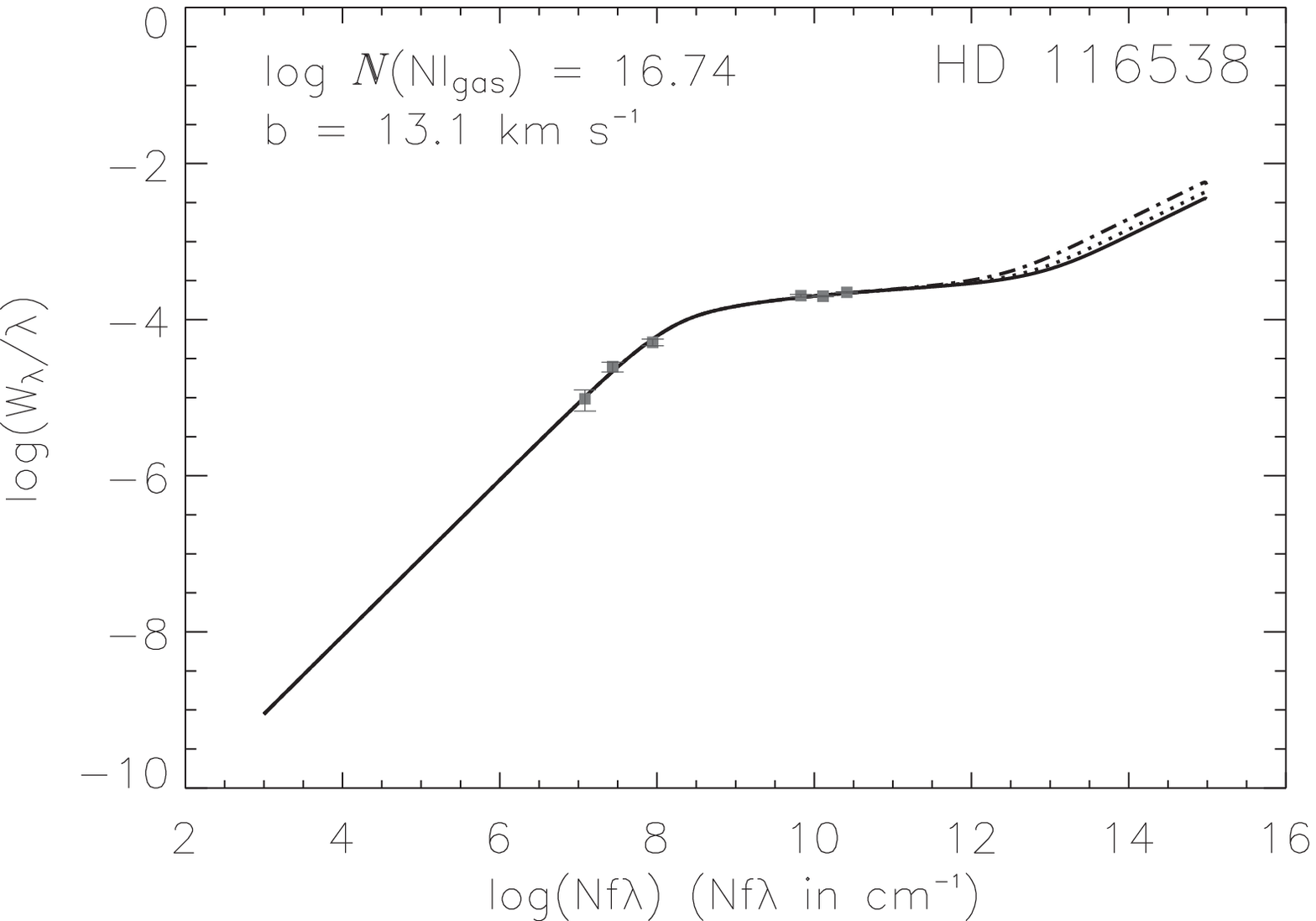}}

\vspace{.3 cm}

\scalebox{.5}[.5]{\includegraphics{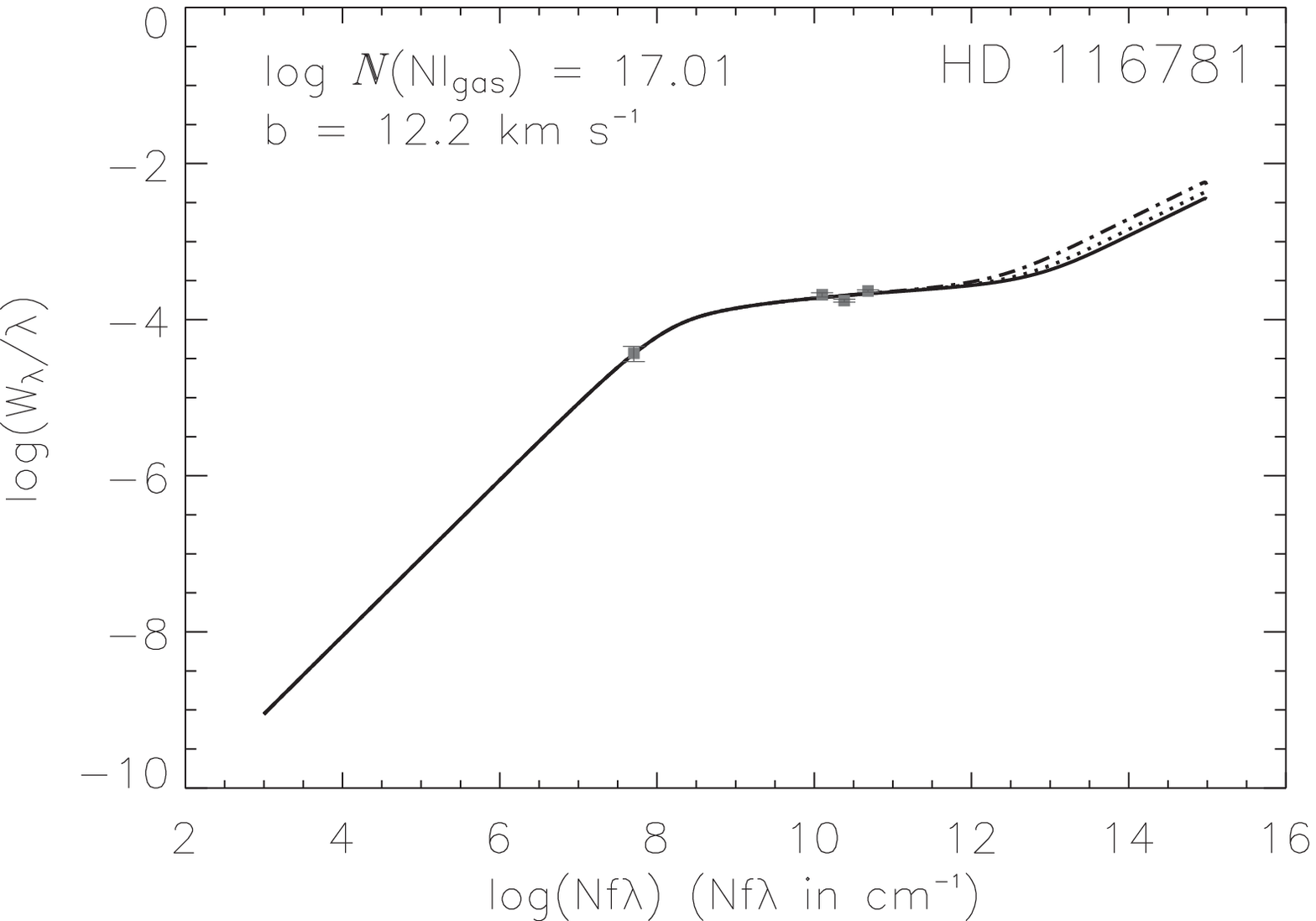} \quad \includegraphics{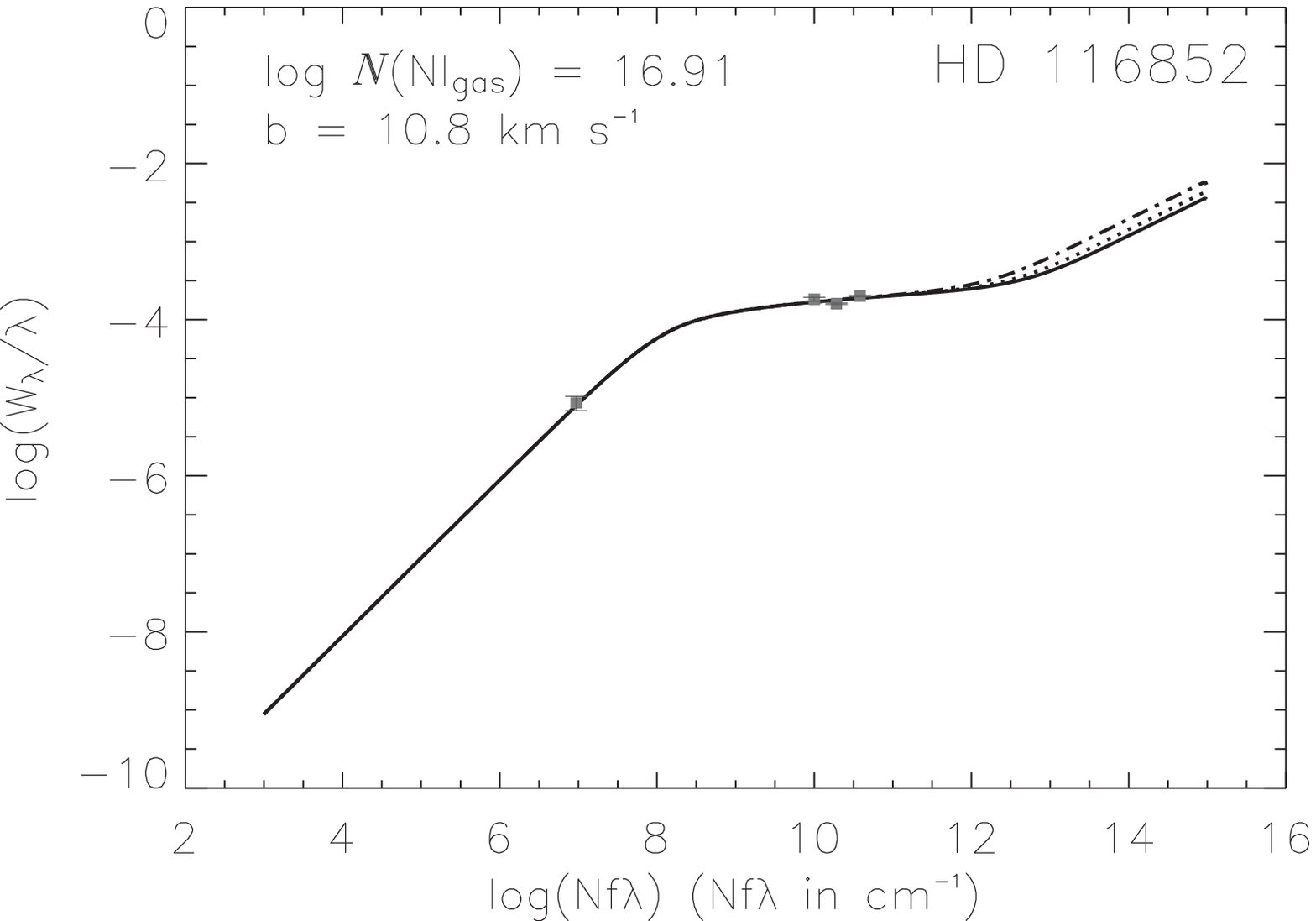}}

\vspace{.3 cm}

\scalebox{.5}[.5]{\includegraphics{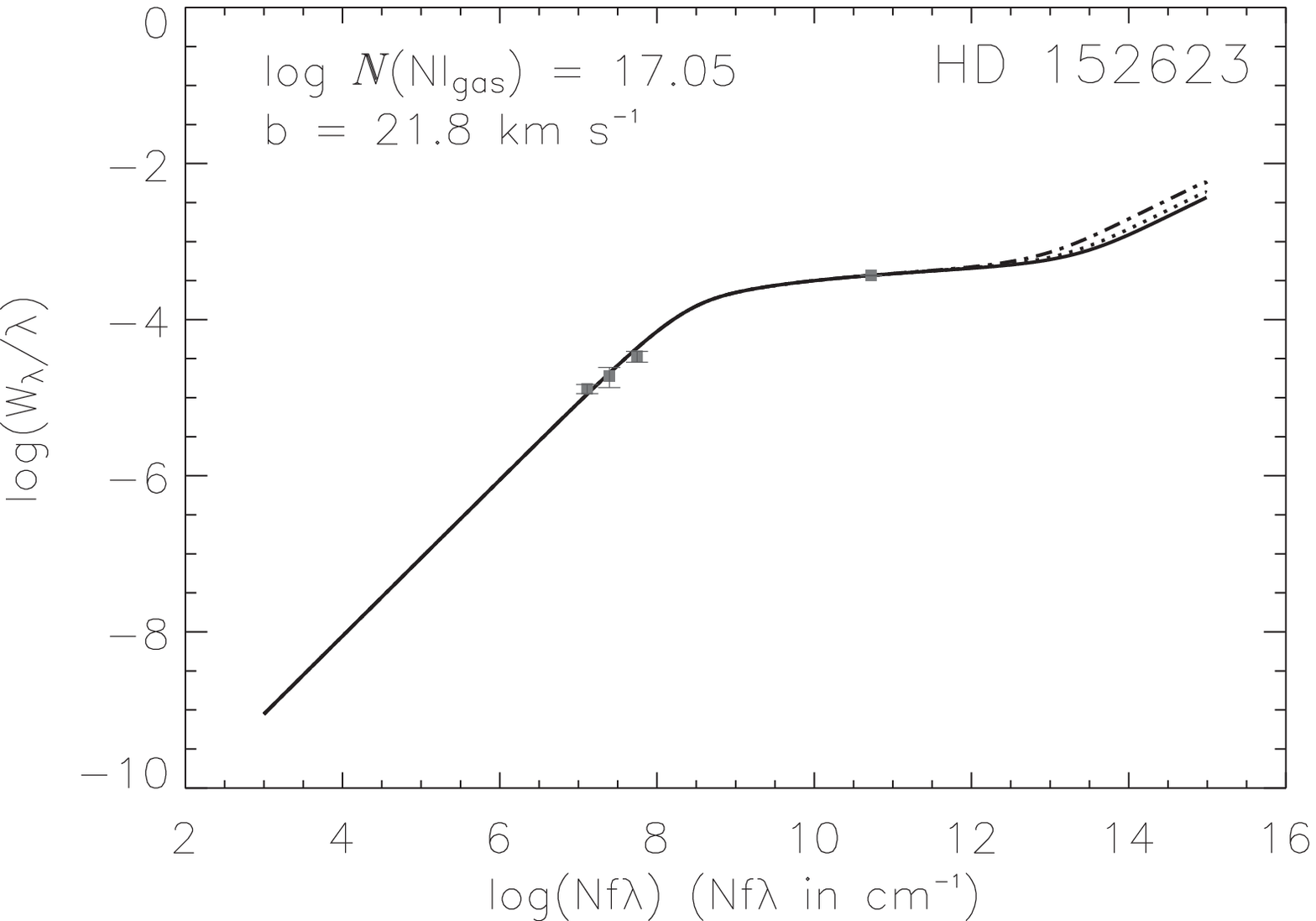} \quad \includegraphics{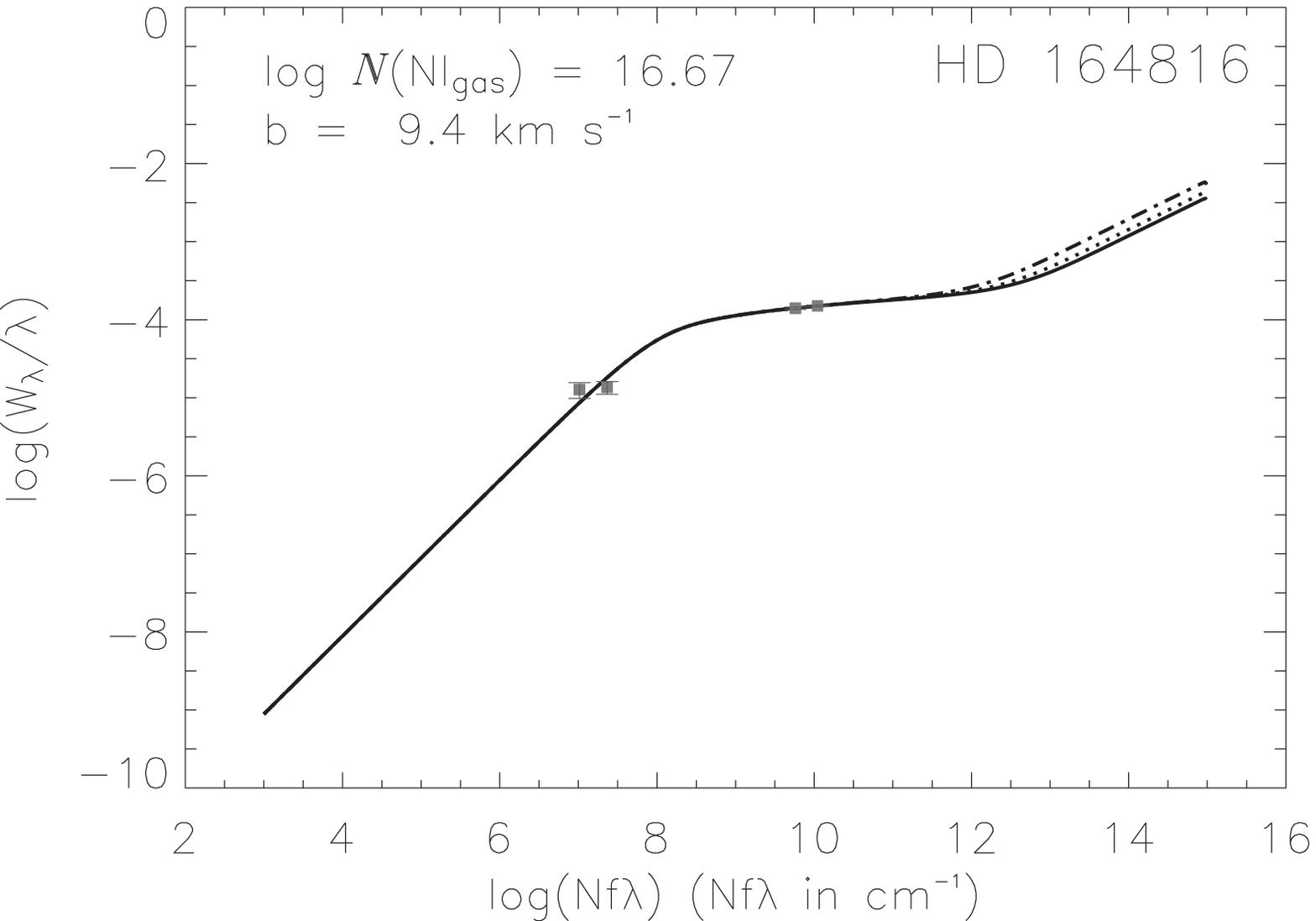}}

\caption{N I curves of growth for HD 103779 through HD 164816.  Symbols and curves are the same as those in Figure \ref{fig_Ncog1}. \label{fig_Ncog2}}
\end{center}
\end{figure}
\end{center}

\begin{center}
\begin{figure}
\begin{center}

\scalebox{.5}[.5]{\includegraphics{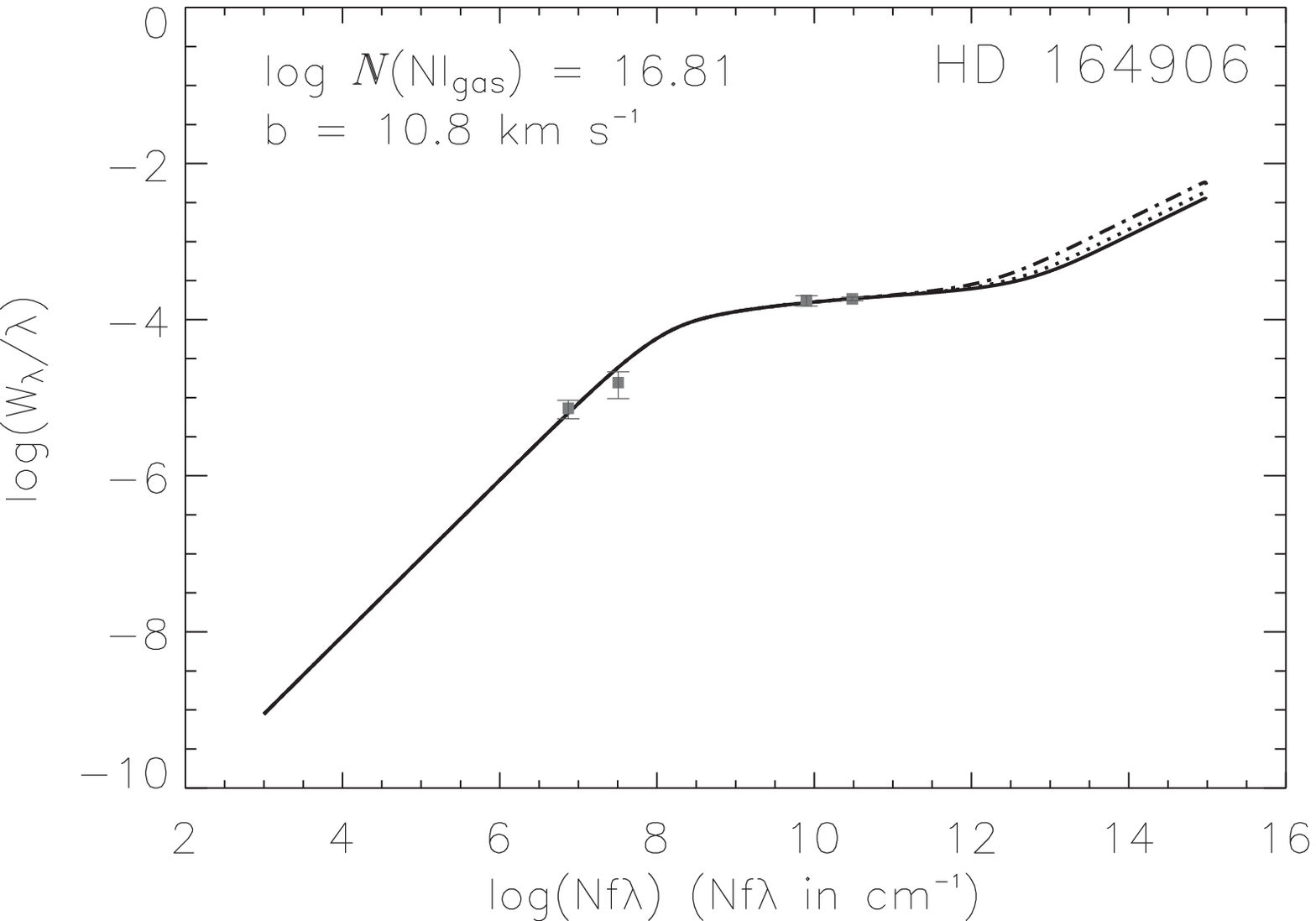} \quad \includegraphics{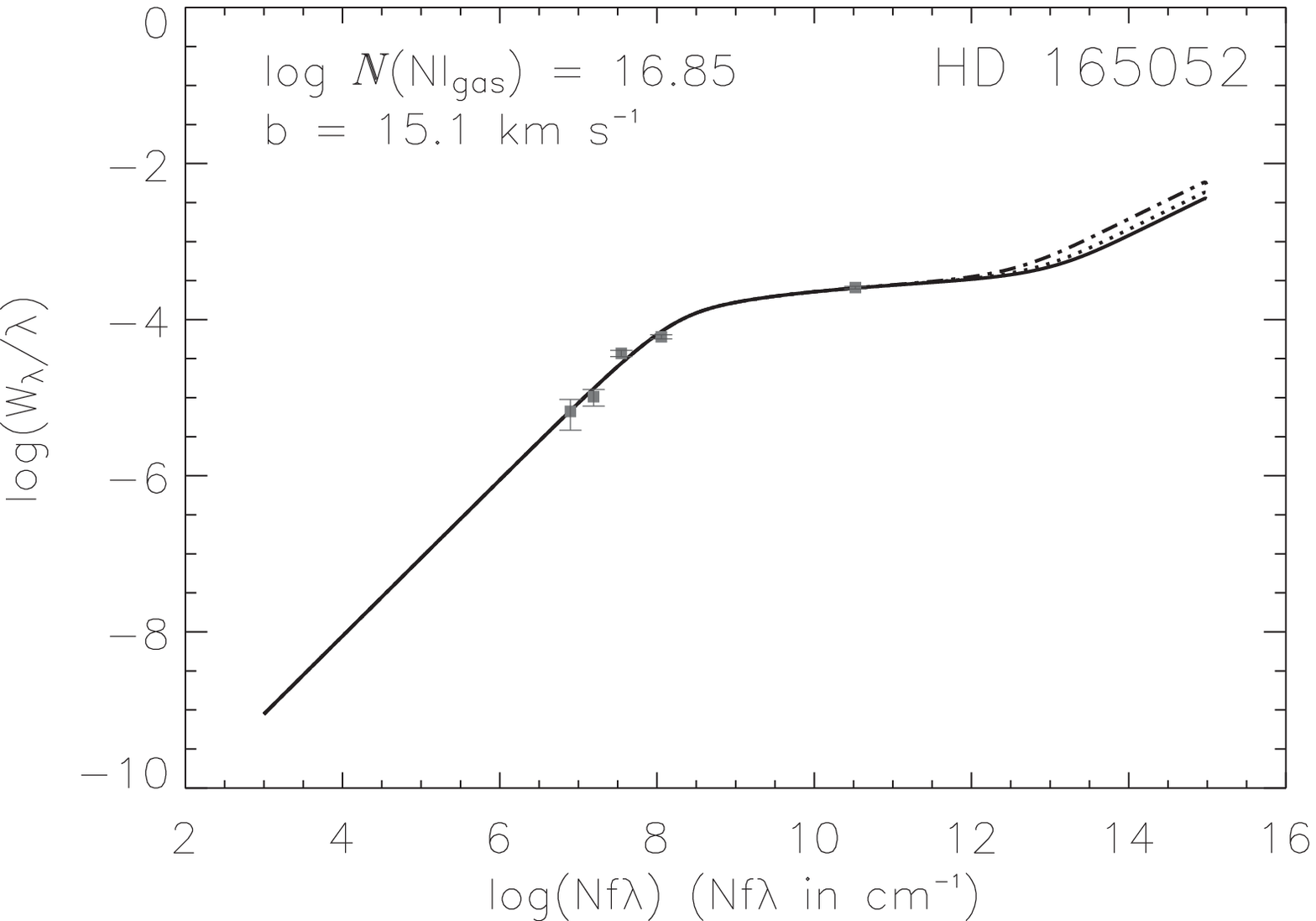}}

\vspace{.3 cm}

\scalebox{.5}[.5]{\includegraphics{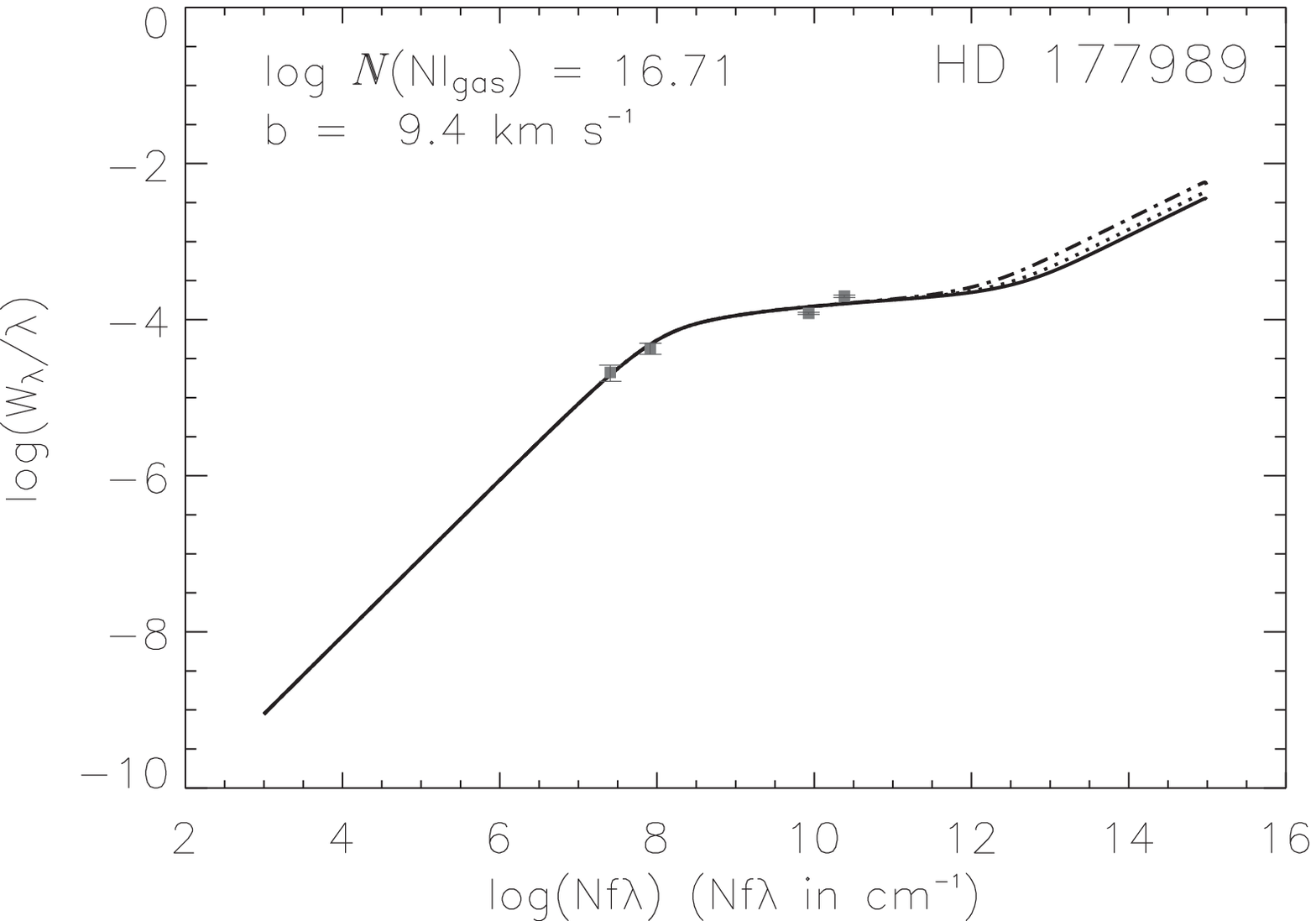} \quad \includegraphics{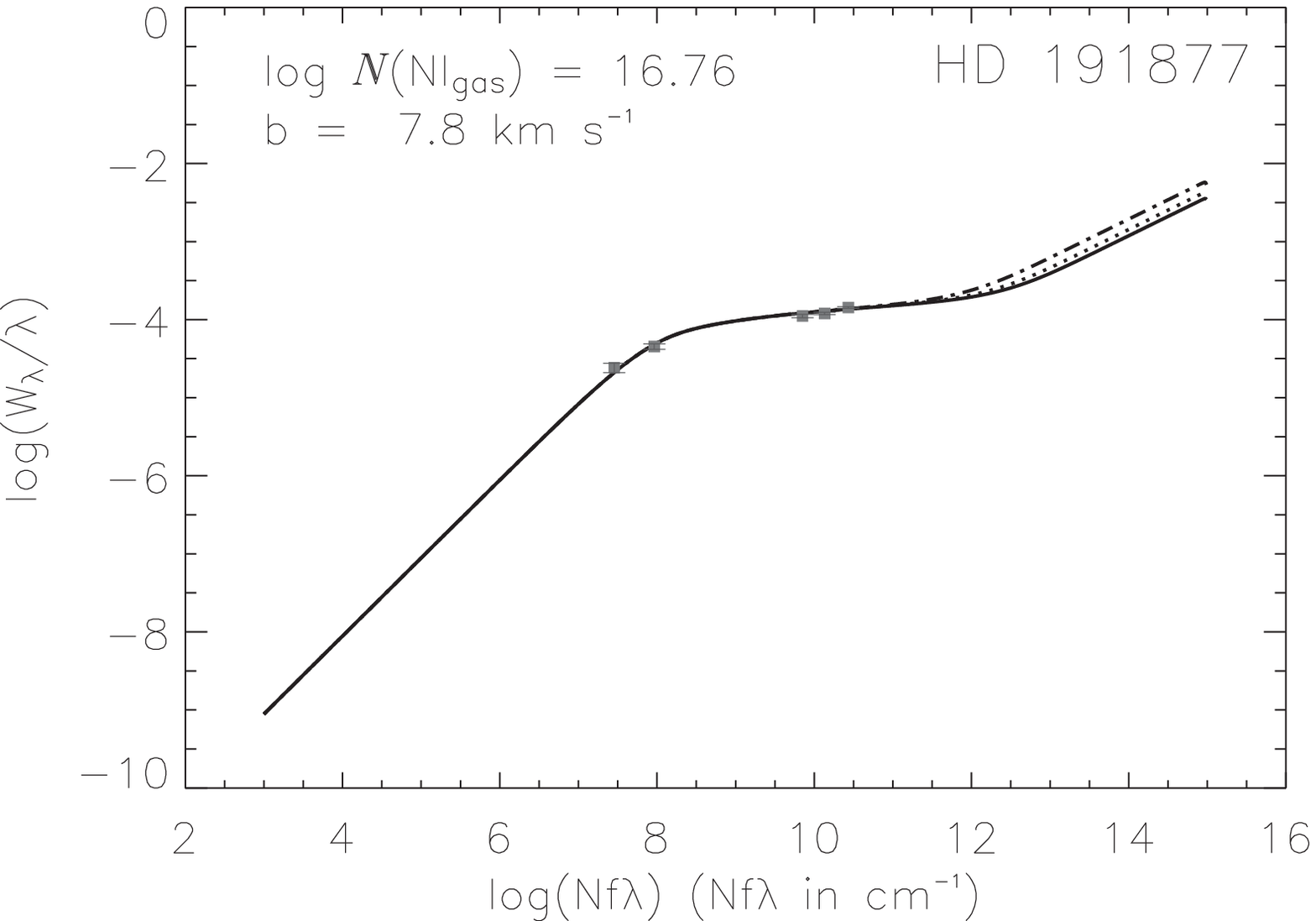}}

\vspace{.3 cm}

\scalebox{.5}[.5]{\includegraphics{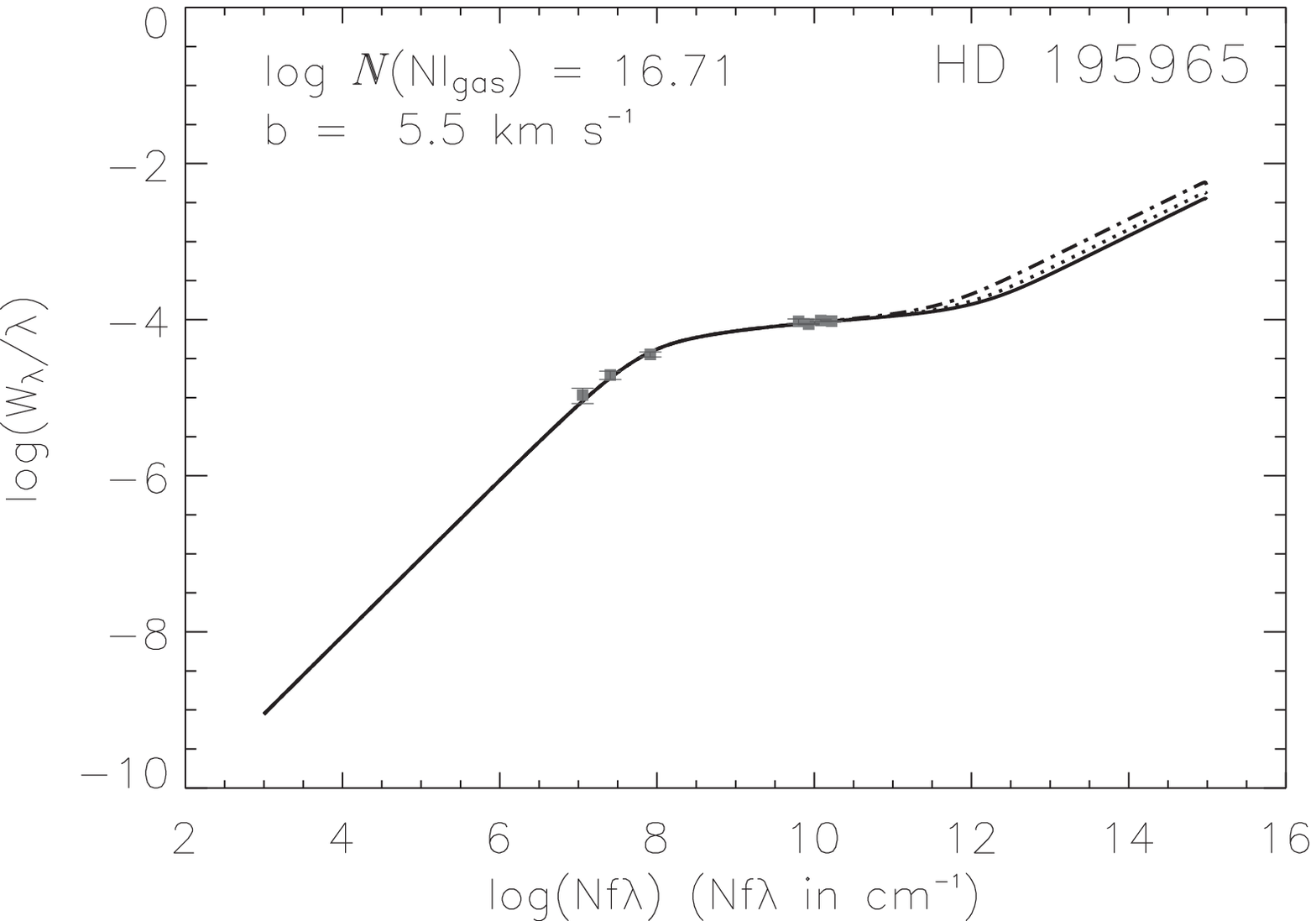} \quad \includegraphics{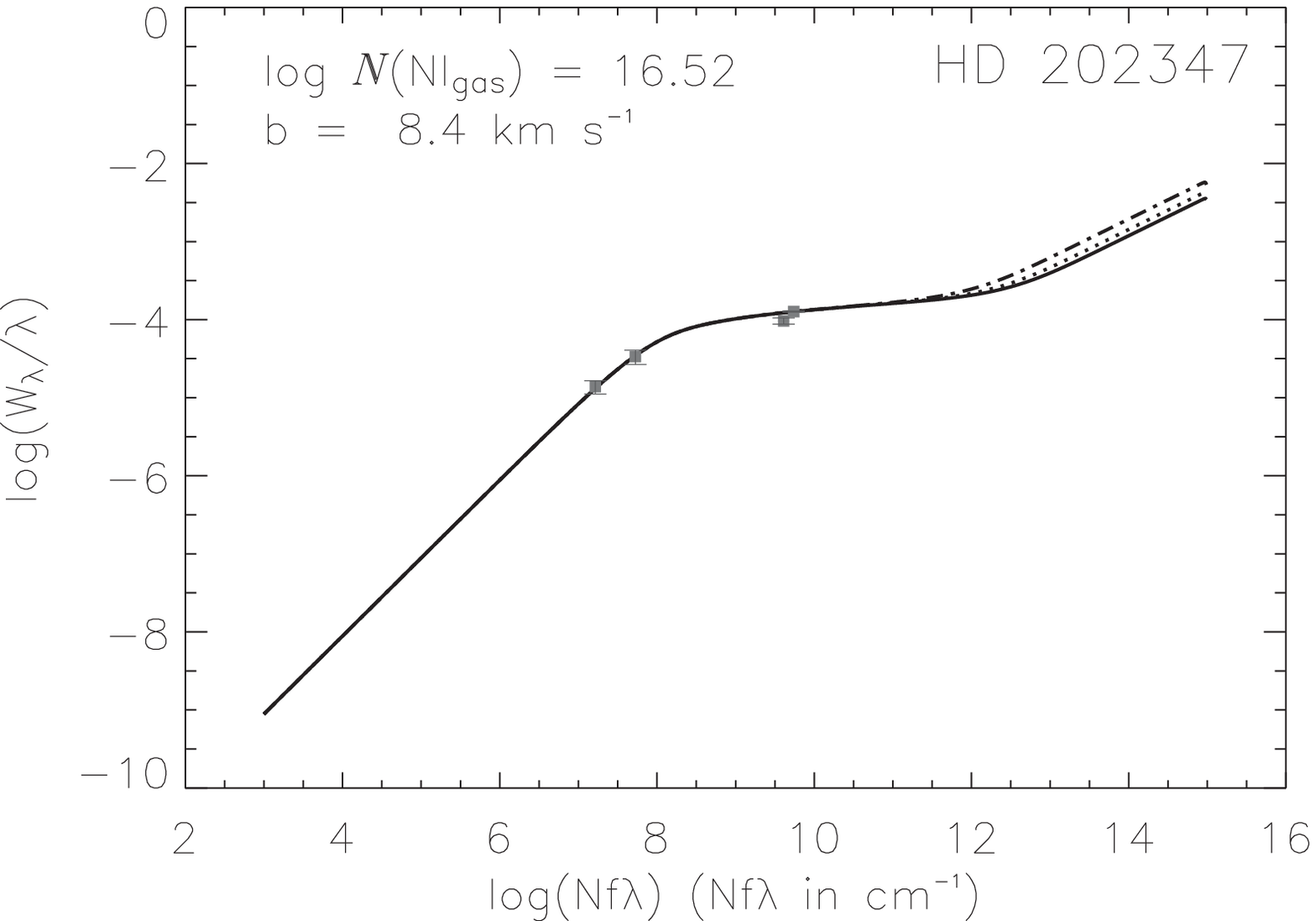}}

\caption{N I curves of growth for HD 164906 through HD 202347.  Symbols and curves are the same as those in Figure \ref{fig_Ncog1}.\label{fig_Ncog3}}
\end{center}
\end{figure}
\end{center}

\begin{center}
\begin{figure}
\begin{center}

\scalebox{.5}[.5]{\includegraphics{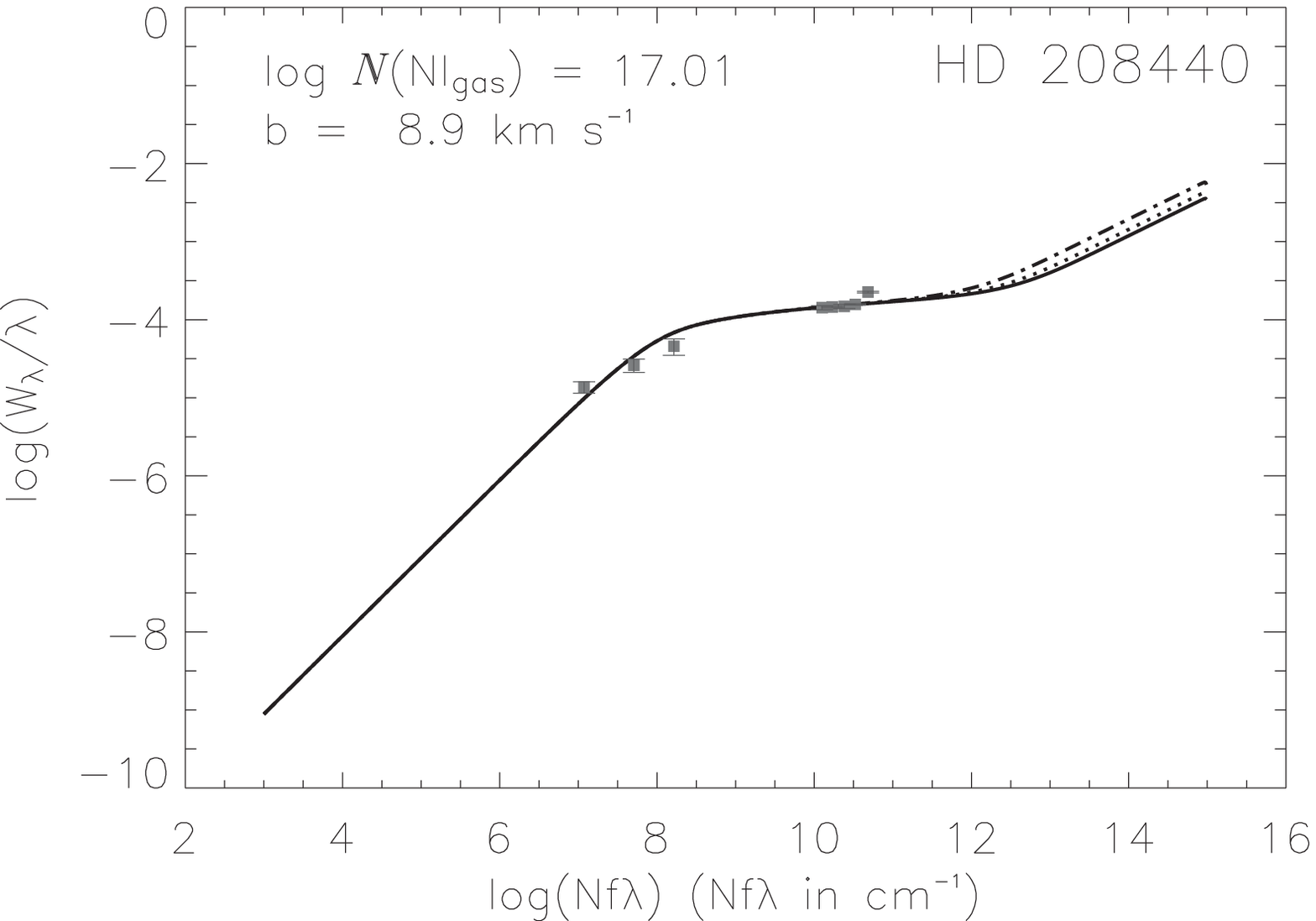} \quad \includegraphics{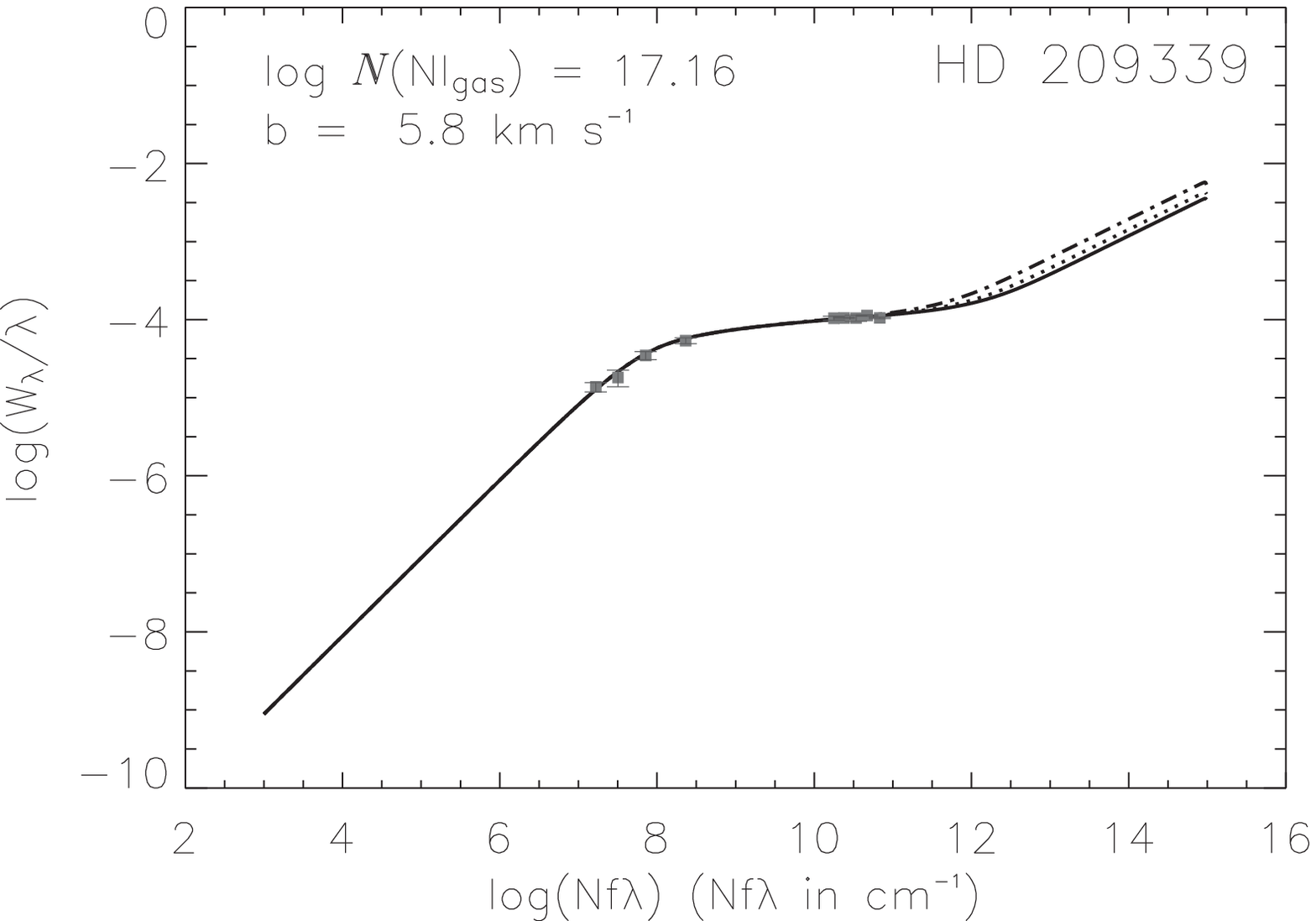}}

\vspace{.3 cm}

\scalebox{.5}[.5]{\includegraphics{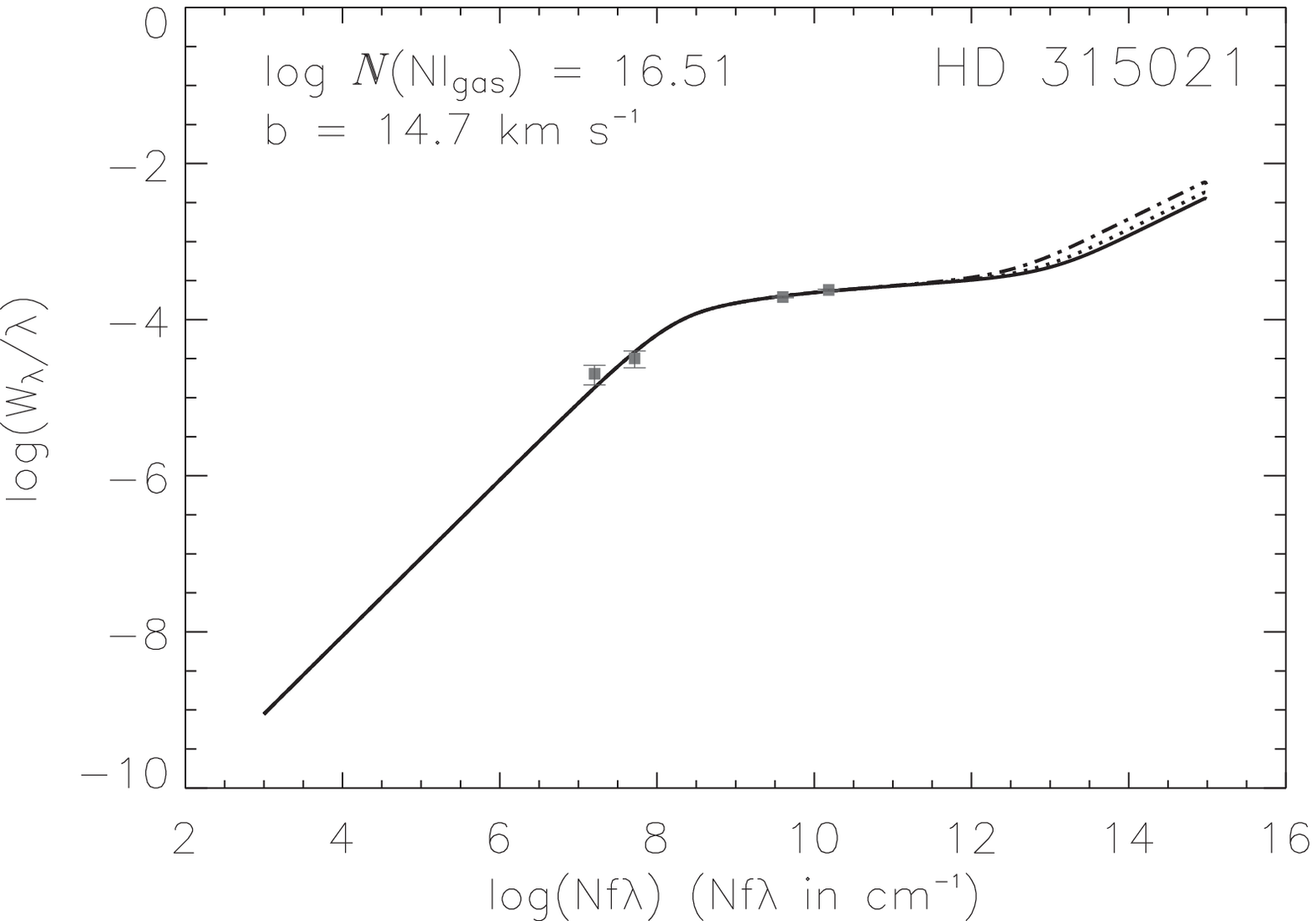}}

\caption{N I curves of growth for HD 208440, HD 209339, and HD 315021.  Symbols and curves are the same as those in Figure \ref{fig_Ncog1}.\label{fig_Ncog4}}
\end{center}
\end{figure}
\end{center}

\begin{center}
\begin{figure}
\begin{center}
\scalebox{.5}[.5]{\includegraphics{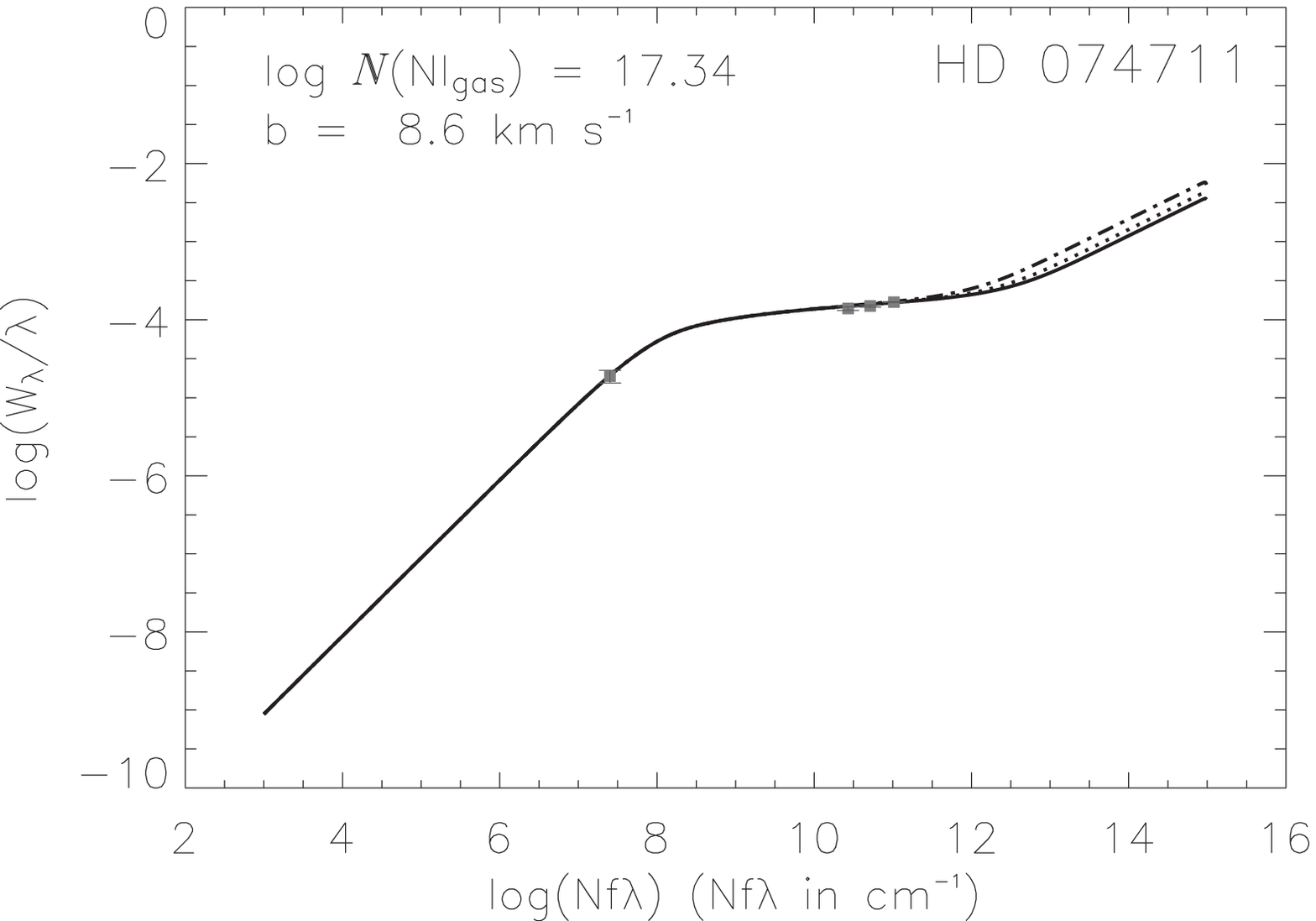} \quad \includegraphics{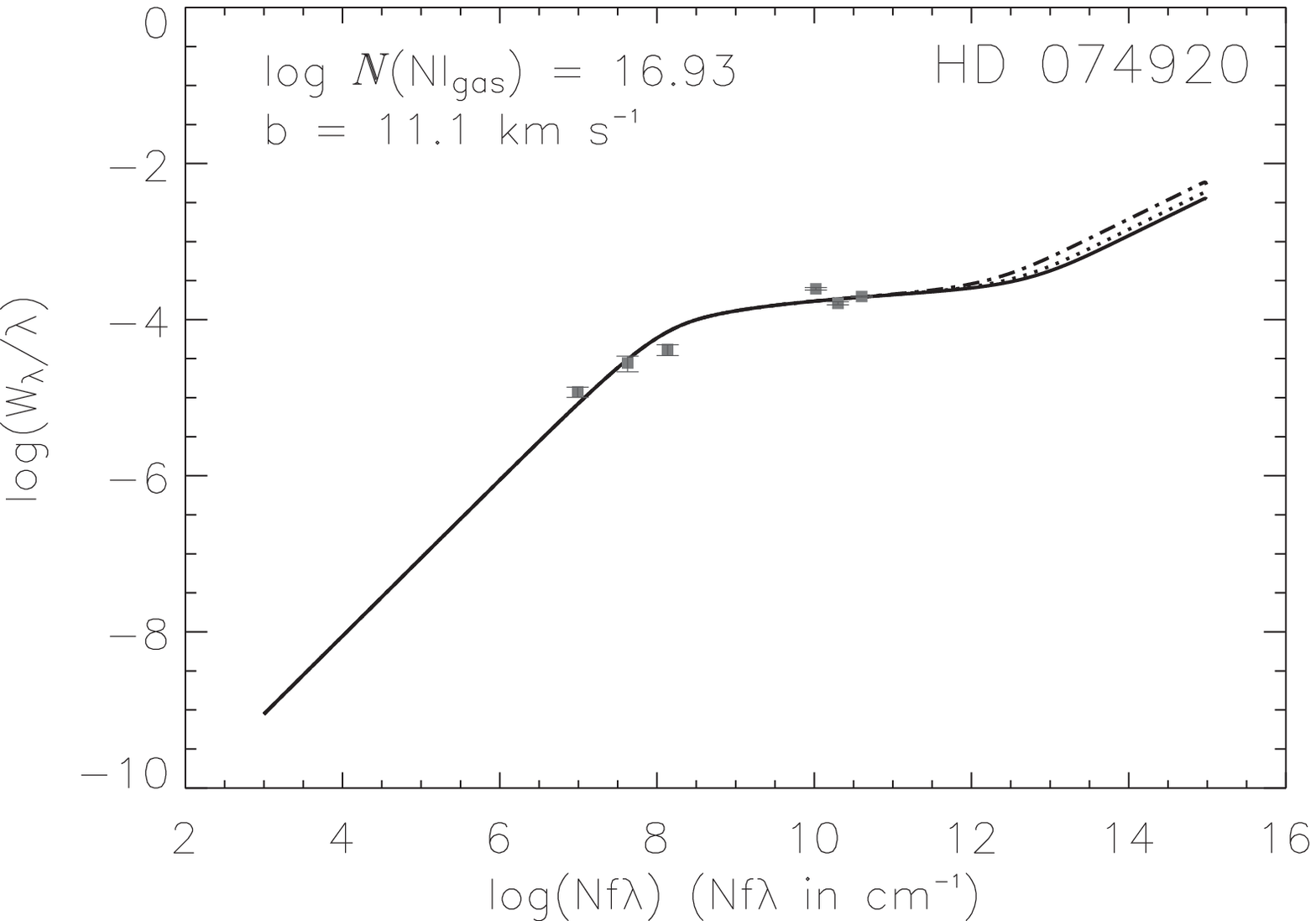}}

\vspace{.3 cm}

\scalebox{.5}[.5]{\includegraphics{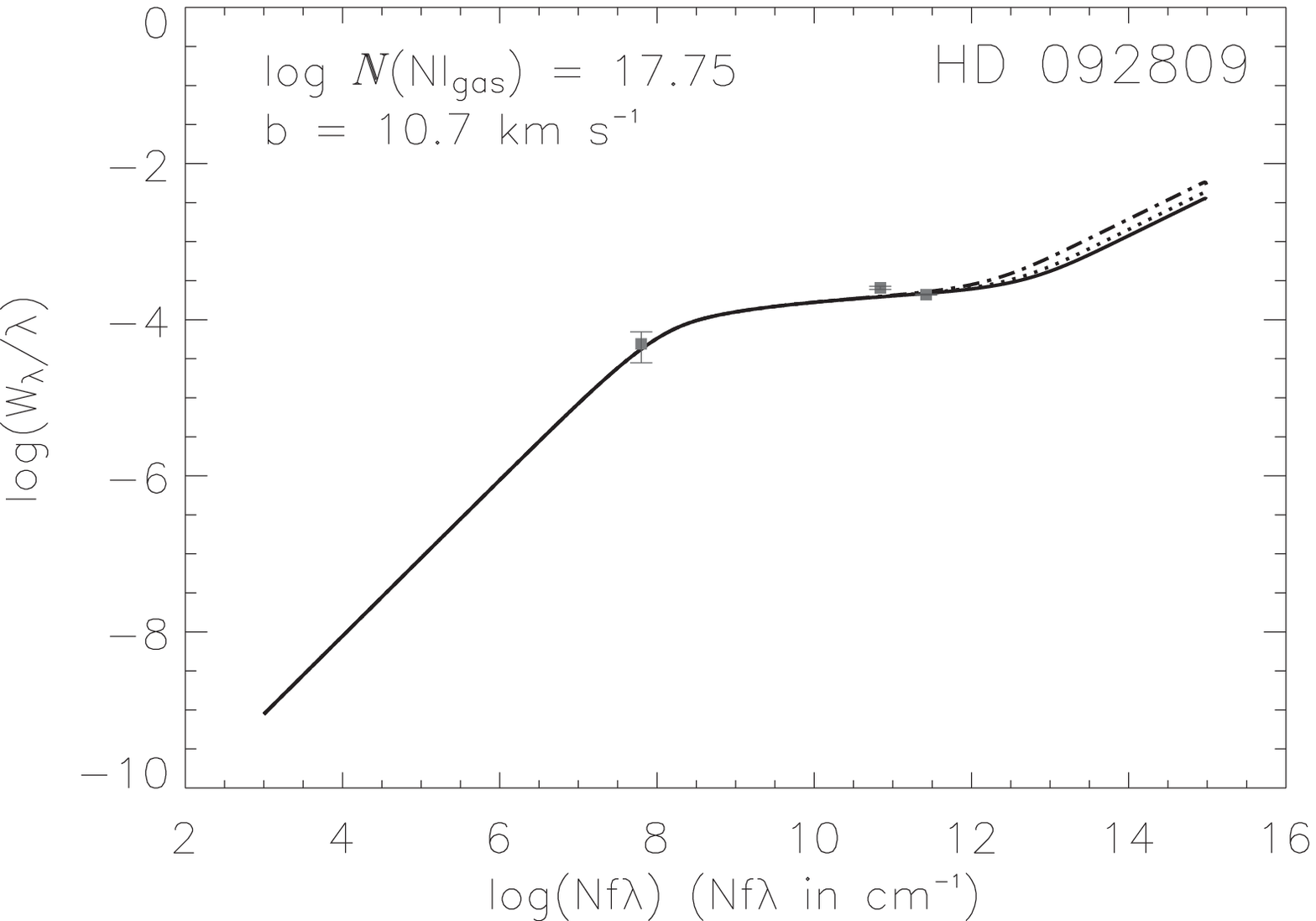} \quad \includegraphics{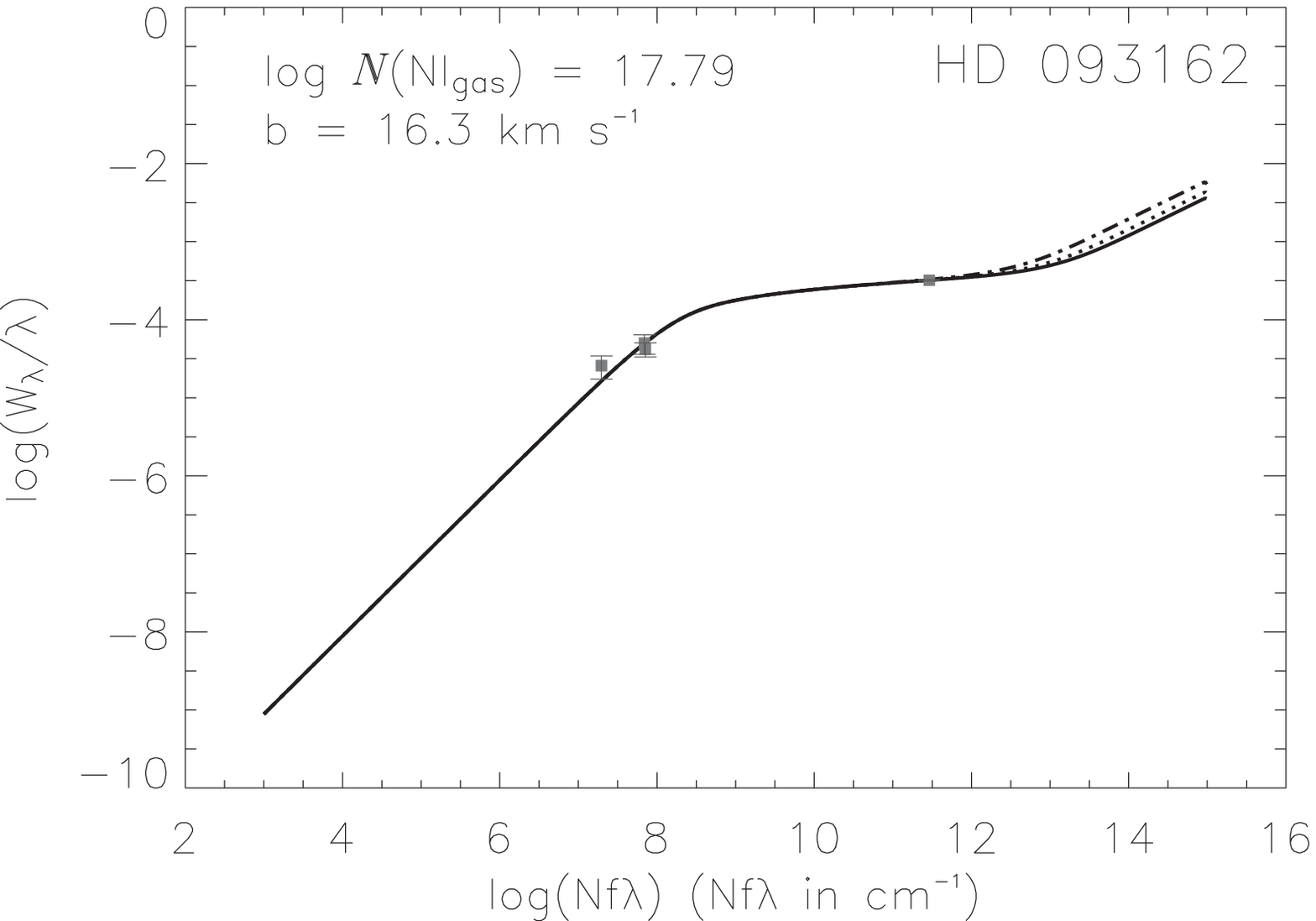}}

\vspace{.3 cm}

\scalebox{.5}[.5]{\includegraphics{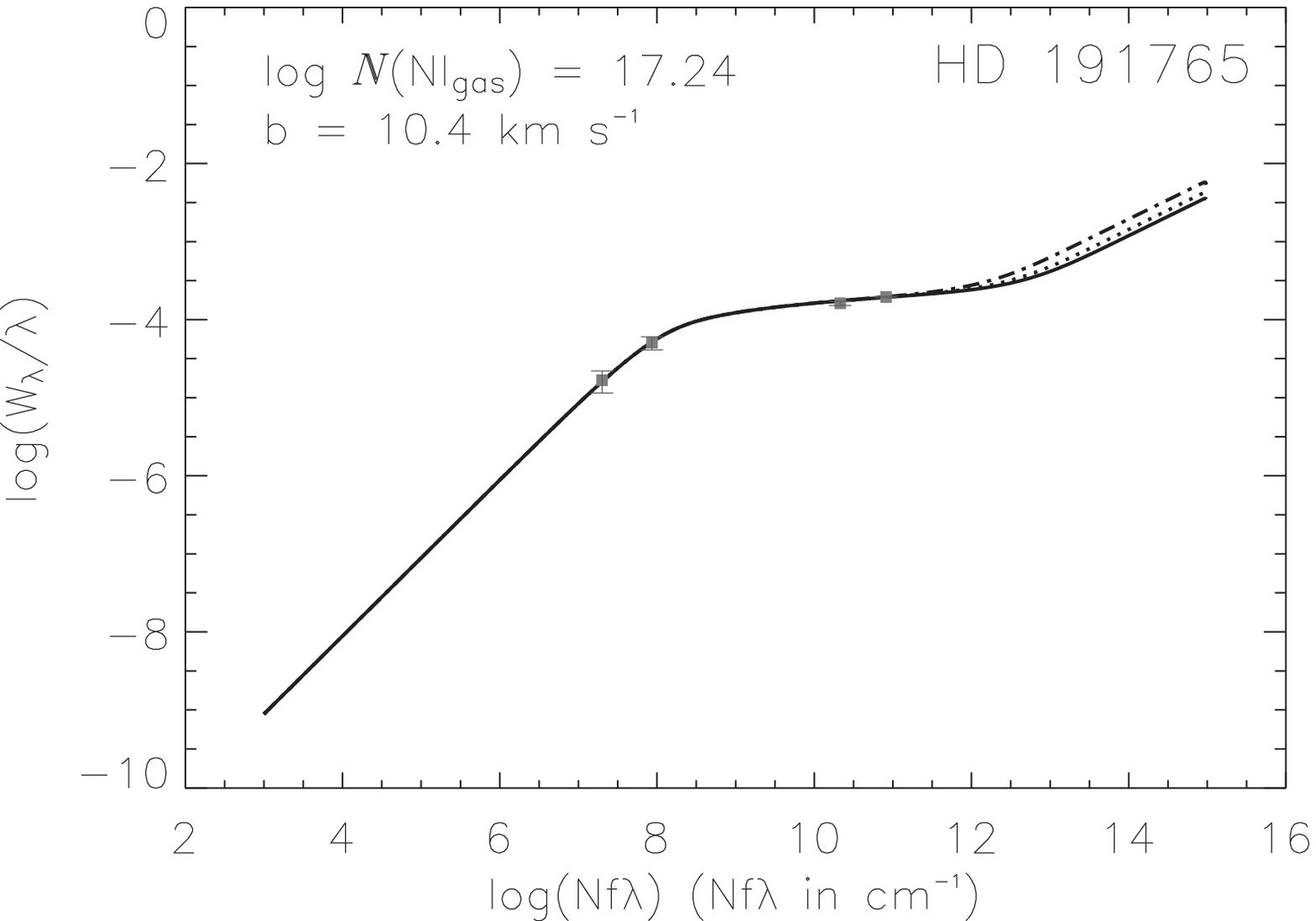}}

\caption{N I curves of growth for lines of sight through supernova remnants and to Wolf-Rayet stars.  Symbols and curves are the same as those in Figure \ref{fig_Ncog1}.\label{fig_Ncog5}}
\end{center}
\end{figure}
\end{center}

To find a curve of growth for N I, we measured the equivalent 
widths for as many N I absorption lines as possible 
within the range of the {\it FUSE} spectrum.  We then fit all these 
measured lines to a curve of growth to find the column density and 
{\it b}-value for N I, assuming a single velocity 
component (see Figures \ref{fig_Ncog1}, \ref{fig_Ncog2}, \ref{fig_Ncog3}, 
\ref{fig_Ncog4}, and \ref{fig_Ncog5}).  Where data were available, we 
checked our assumption of a single velocity component by observing the 
1356 \AA\ O I line in {\it HST} STIS spectra.  For most of the lines 
of sight that had {\it HST} data, we found only a single velocity component.  
Two sightlines (HD 37367 and HD 93250) showed two or more clearly 
separated velocity components.  However, we find that in these cases 
the N I curves of growth are still internally consistent.  Furthermore, 
the broad separation of the components of the 1356 \AA\ O I line 
indicates that the total profiles of lines in FUSE data should
still be optically thin as long as each component is optically thin.
Again, this is supported by the internal consistency of the curves of
growth, even when a large {\it b}-value is implied.  The F I lines under
examination are approximately as weak as or weaker than the weakest N I
lines used in the curves of growth, and the use of the derived N I
{\it b}-values to place limits on saturation is therefore justified.
N I column densities and {\it b}-values are given in Table \ref{tab_NColDen}.  
For more details on the procedure used for deriving the N I curve of 
growth see Jensen, Rachford, \& Snow, (2006).

\subsection{Fluorine Column Densities}
Once a curve of growth was established for N I, we used this same 
curve to find the column densities for F I from our measured 
equivalent widths.  Errors in column densities were derived by 
carrying through the one-sigma errors on the Gaussian fit of the 
F I line.  Errors in the curve of growth itself are not 
significant because the F I lines are weak enough in most cases to be 
on the linear part of the curve.  In cases where we 
could not obtain a significant measurement of the fluorine equivalent 
width, we calculated a 2$\sigma$ upper limit for the column density. 
Final fluorine column densities and limits are given in Table 
\ref{tab_FColDen}.

\subsection{Comparison to Previous Work}
The only other studies deriving interstellar F I column 
densities were done by Snow and York (1981) and by Federman et al. (2005).

Snow and York, using data from the {\it Copernicus} satellite, 
detected the 954 \AA\ F I line toward $\delta$ Scorpii, and derived an 
F/H ratio based on the assumption that the F I line was weak enough 
to be on the linear portion of the curve of growth.  They found an F I 
column density of $13.18^{\ 0.15}_{-0.14}$.

The Federman et al. study based on \FUSE\ spectra analyzed 
two lines of sight, which we have also included in our survey: 
HD 208440 and HD 209339.  A comparison of results can be found in Table 
\ref{tab_Federmancompare}.  We find that our results match those from 
the Federman et al. study almost exactly, well withing the 1-$\sigma$ errors.

\subsection{The 951 \AA\ Line of F I}

The secondary line of F I lies at 951.871 \AA, which is also 
accessible to {\it FUSE}.  The f-value of this line is weaker by a factor of five 
than the f-value of the 954 \AA\ line, but in some cases might be detectable.  
We may have detected the line in five cases.  Where we did not detect it, 
the upper limit helped us to constrain the curve of growth independent of the N 
I analysis.

\begin{center}
\begin{figure}
\begin{center}
\scalebox{.35}[.35]{\includegraphics[angle=90]{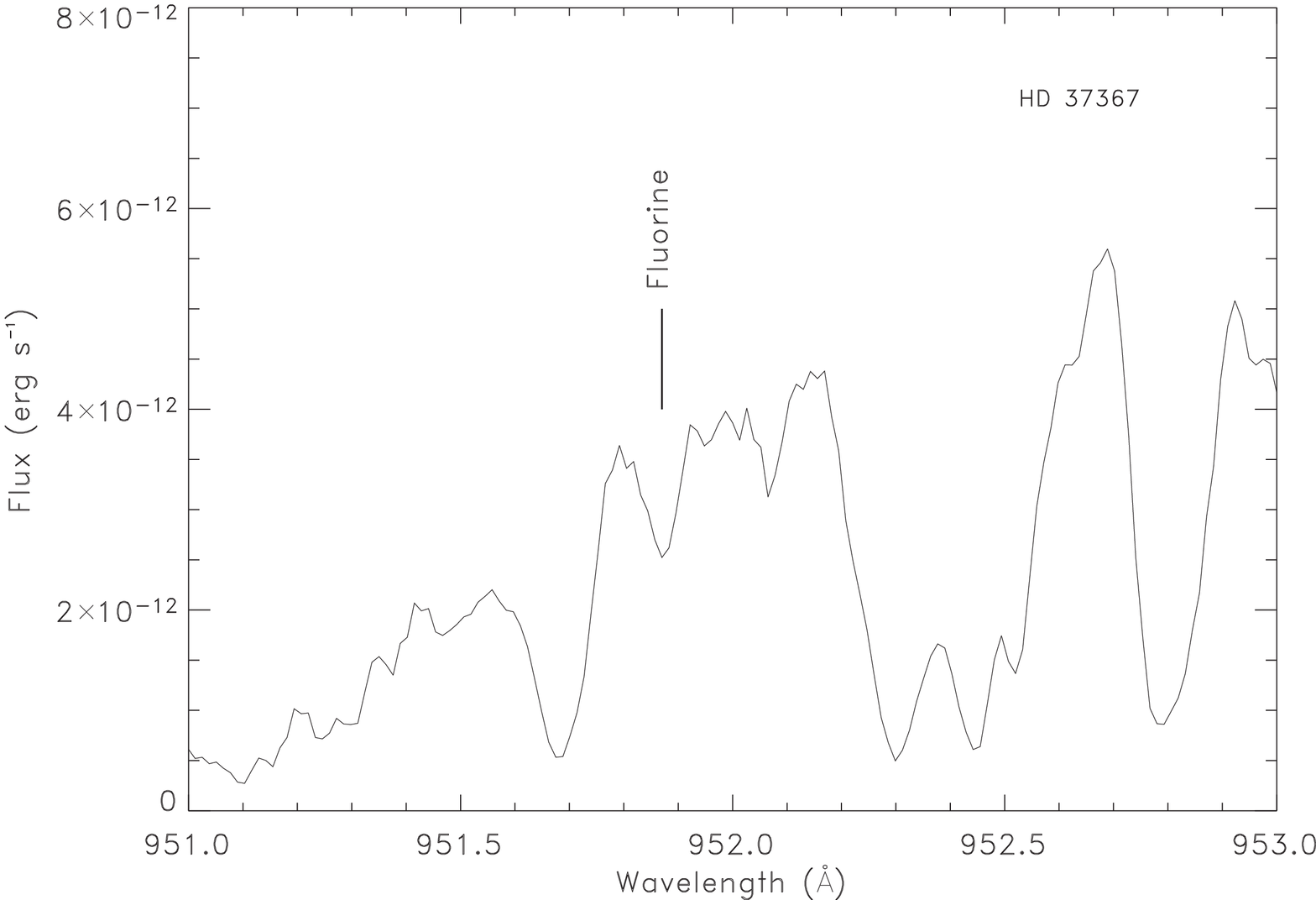}  \quad \includegraphics[angle=90]{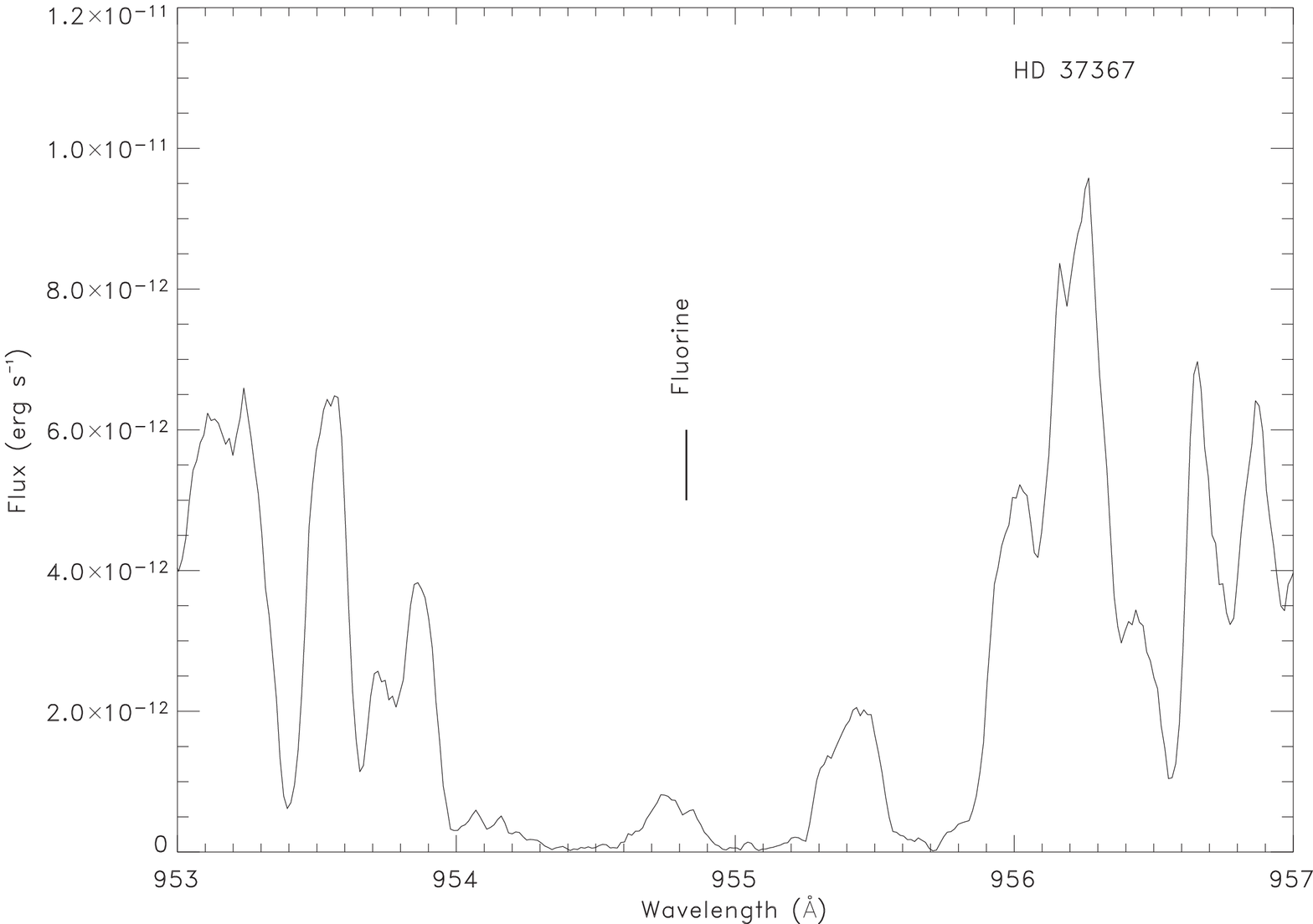}}

\vspace{.3 cm}

\scalebox{.35}[.35]{\includegraphics[angle=90]{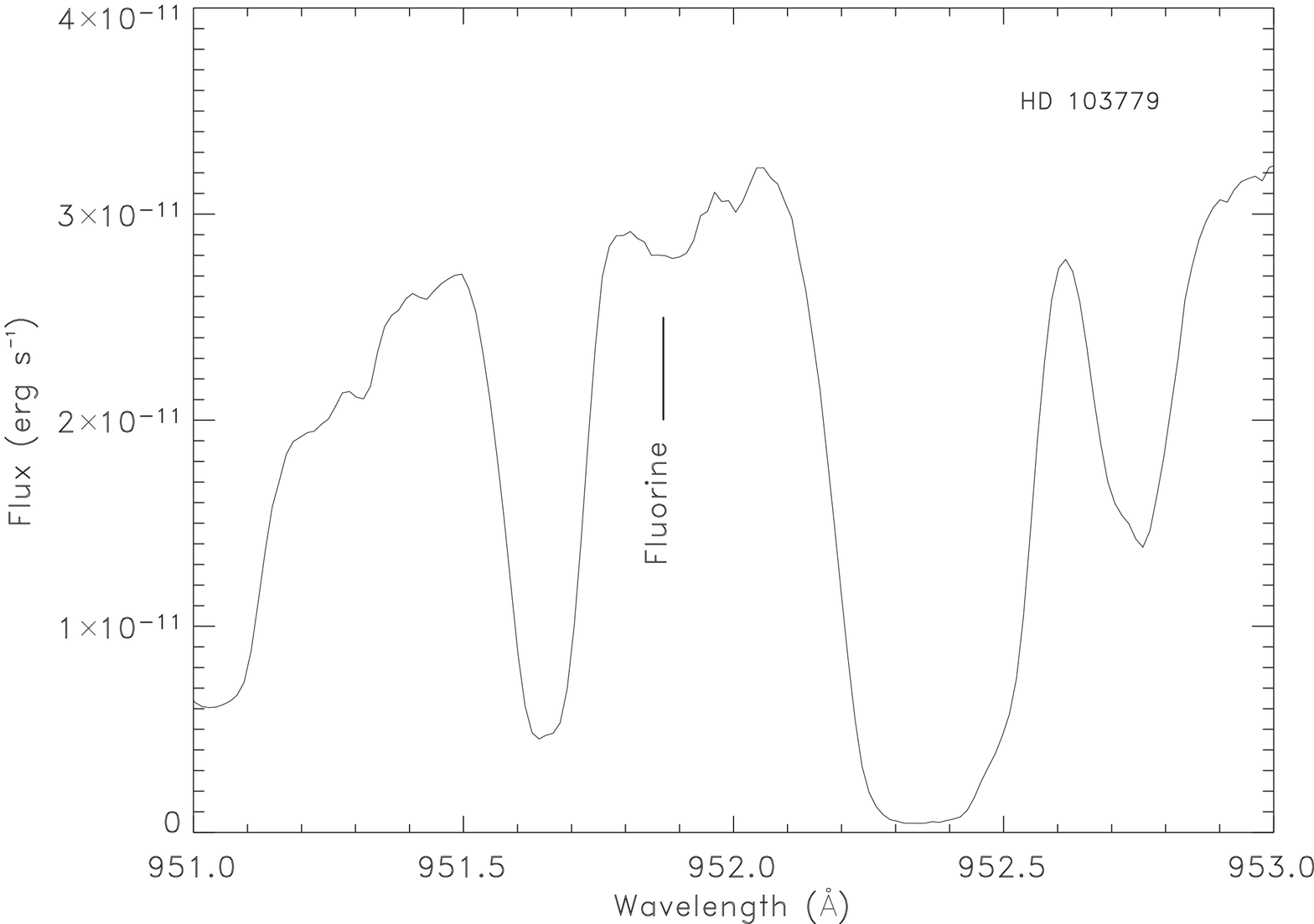} \quad \hspace{.5 cm} \includegraphics[angle=90]{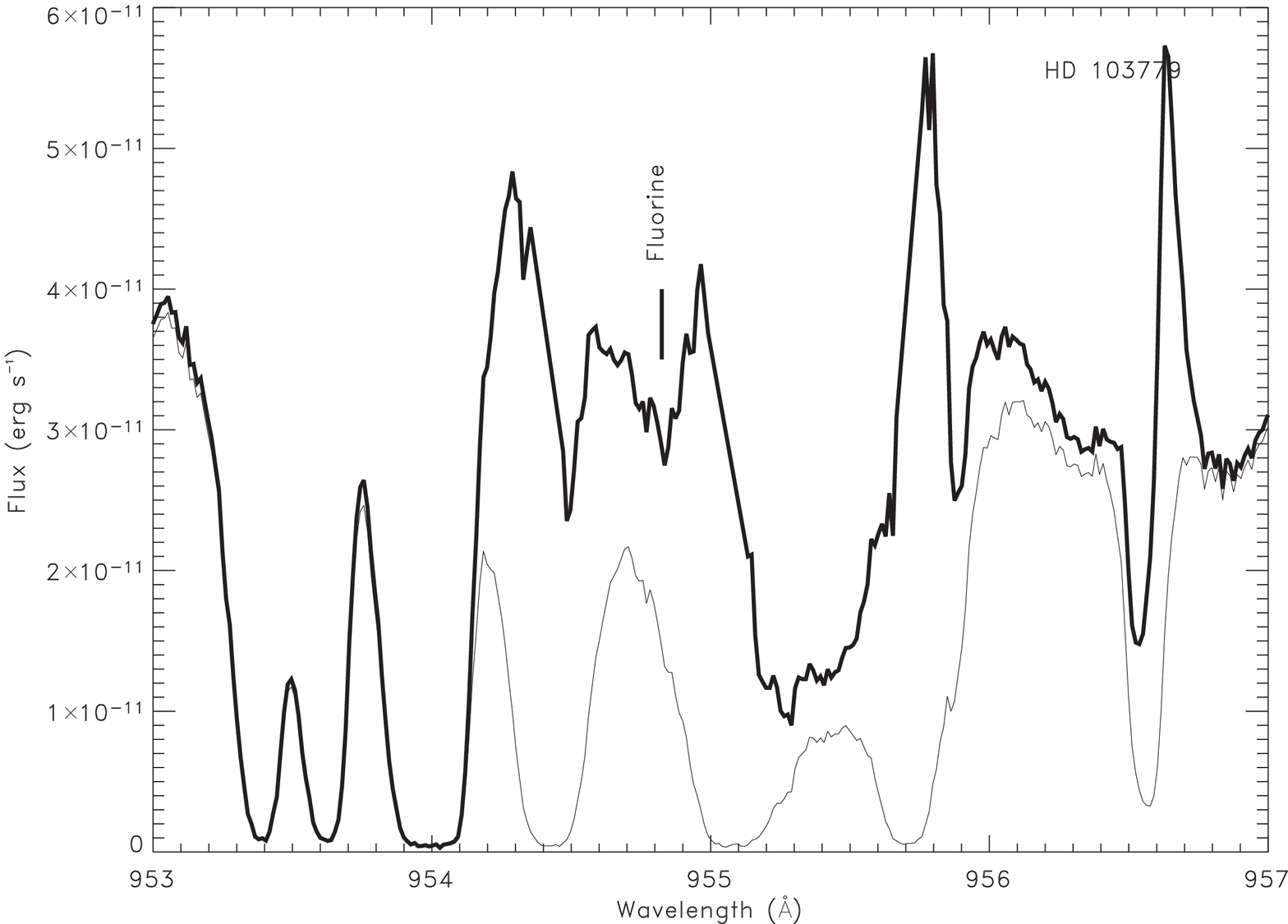}}

\vspace{.3 cm}

\scalebox{.35}[.35]{\includegraphics[angle=90]{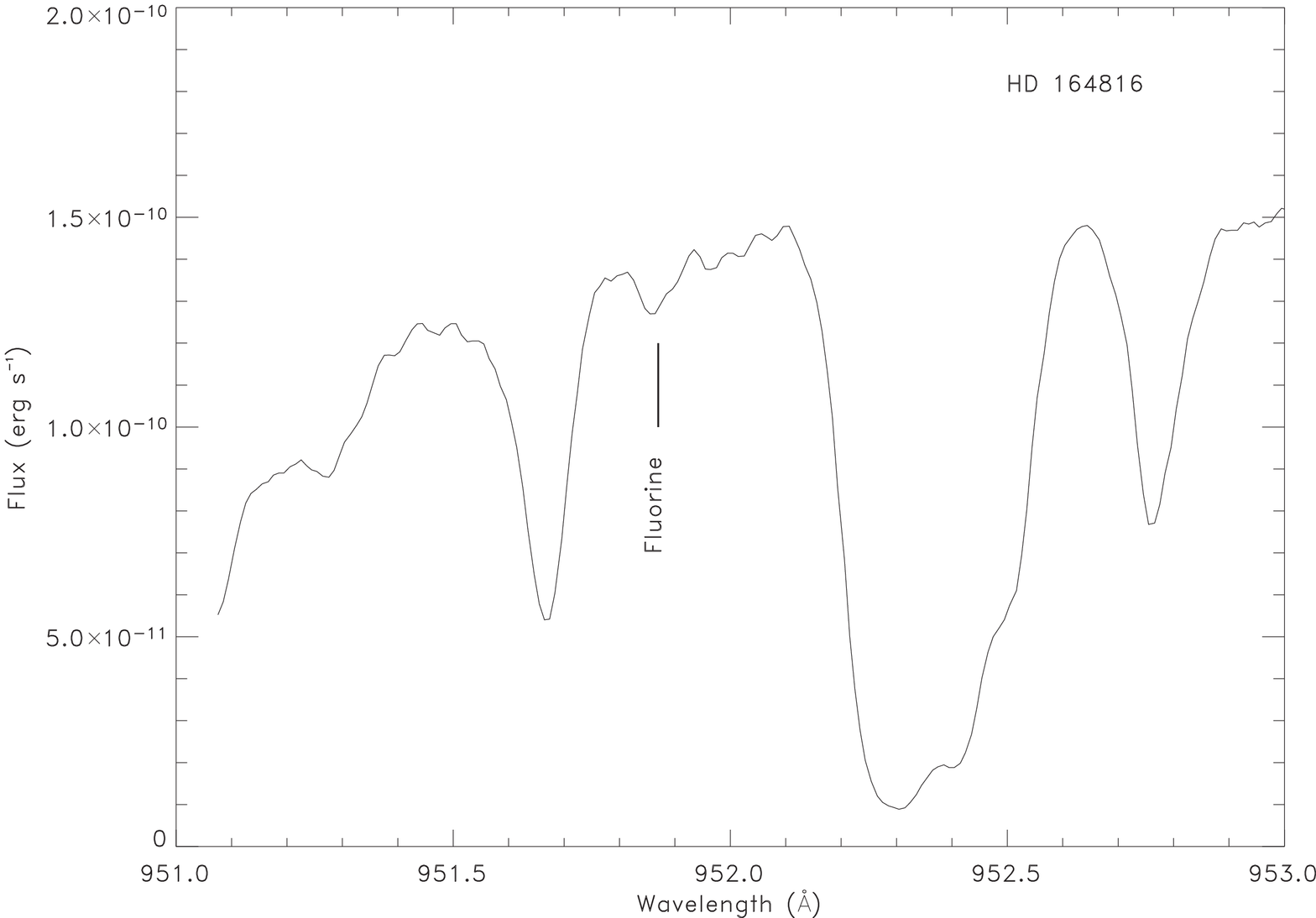} \quad \includegraphics[angle=90]{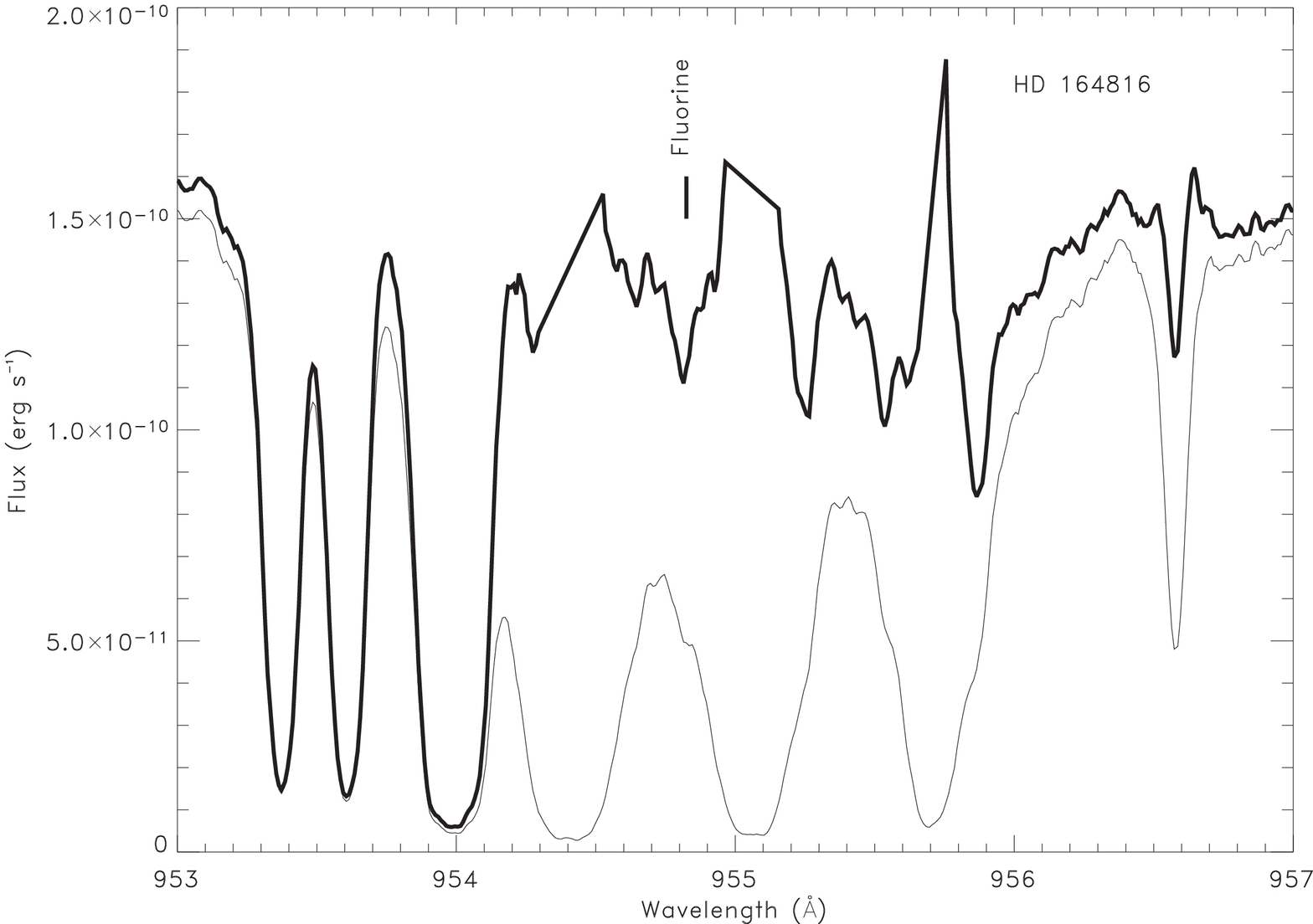}}

\caption{Regions surrounding the 951 \AA\ (left) and 954 \AA\ (right) FI lines for HD 37367 (top), HD 103779 (middle), and HD 164816 (bottom).  Both the original spectra and the spectra after the $H_2$ was divided out (in bold) are shown in the 954 \AA\ case for HD 103779, and HD 164816.\label{fig_954_951_1}}
\end{center}
\end{figure}
\end{center}

\begin{center}
\begin{figure}
\begin{center}

\scalebox{.34}[.34]{\includegraphics[angle=90]{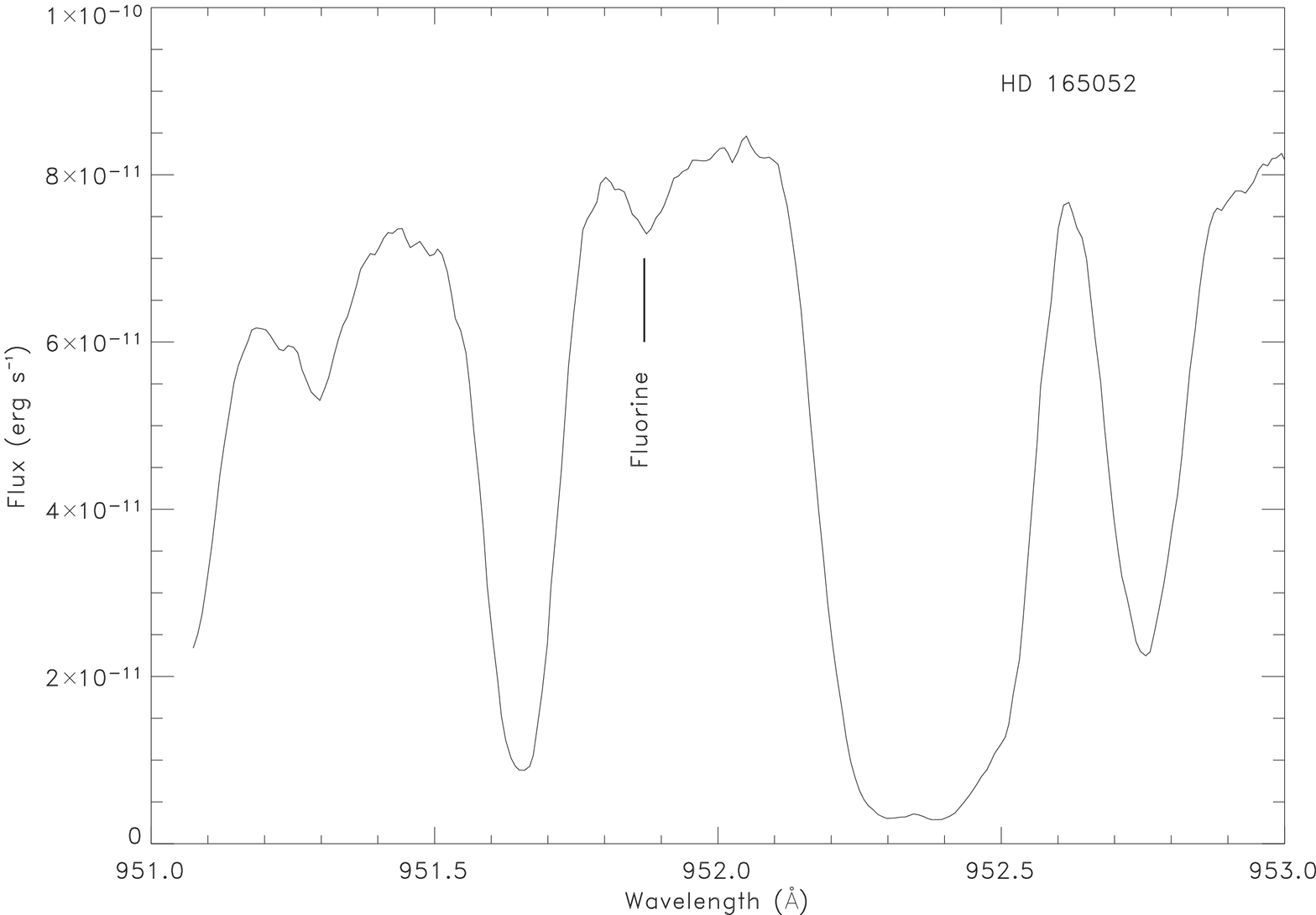} \quad \includegraphics[angle=90]{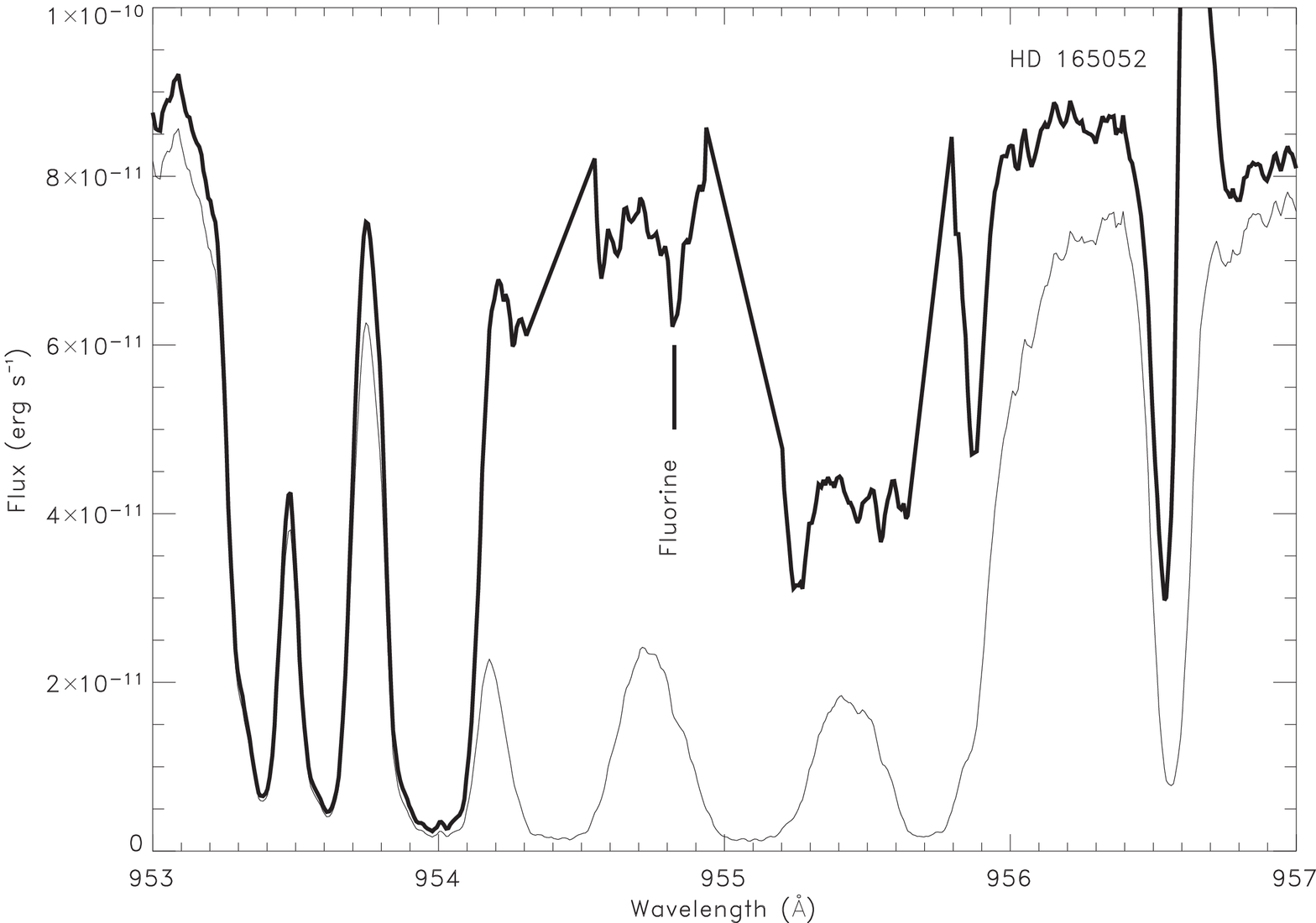}}

\vspace{.3 cm}

\scalebox{.34}[.34]{\includegraphics[angle=90]{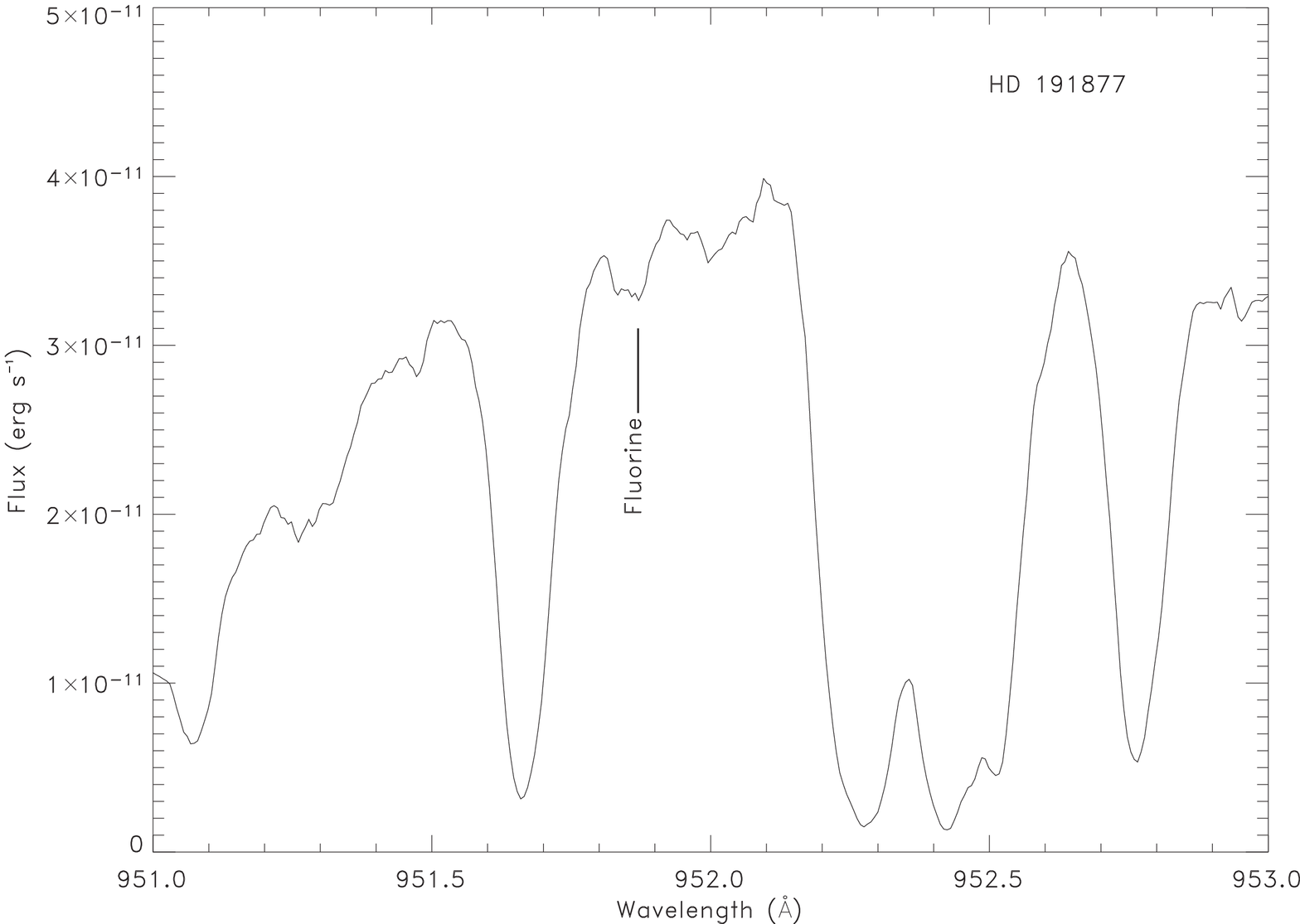} \quad \includegraphics[angle=90]{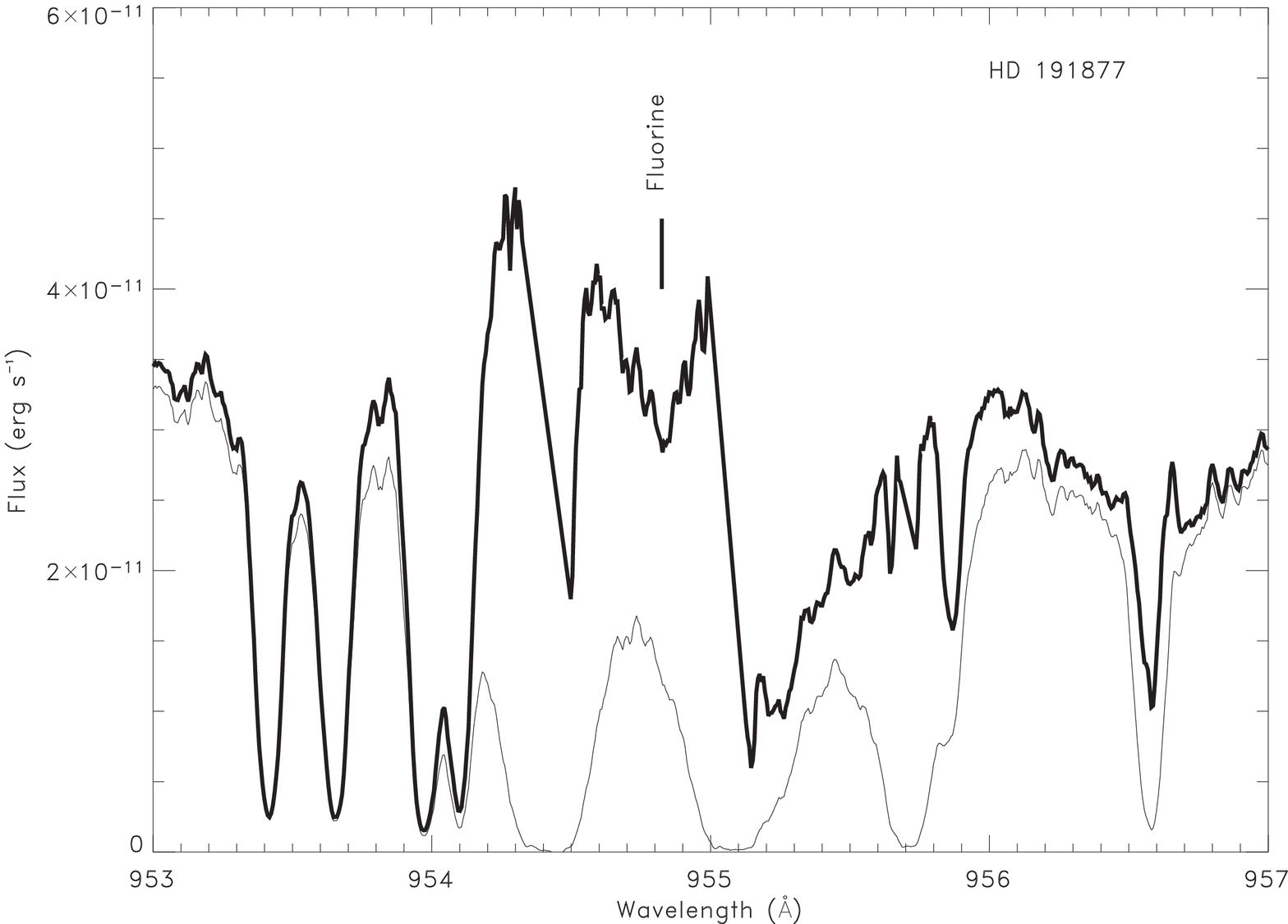}}

\caption{Same as Figure \ref{fig_954_951_1} but for HD 165052 (top) and HD 191877 (bottom).\label{fig_954_951_2}}
\end{center}
\end{figure}
\end{center}

The sightlines where we may have detected the 951 \AA\ F I line are HD 37367, HD 103779, 
HD 164816, HD 165052, and HD 191877.  Plots of both the suspected 951 \AA\ and 954 \AA\ 
lines for these stars are given in Figures \ref{fig_954_951_1} and \ref{fig_954_951_2}.

\begin{center}
\begin{figure}
\begin{center}

\scalebox{.5}[.5]{\includegraphics{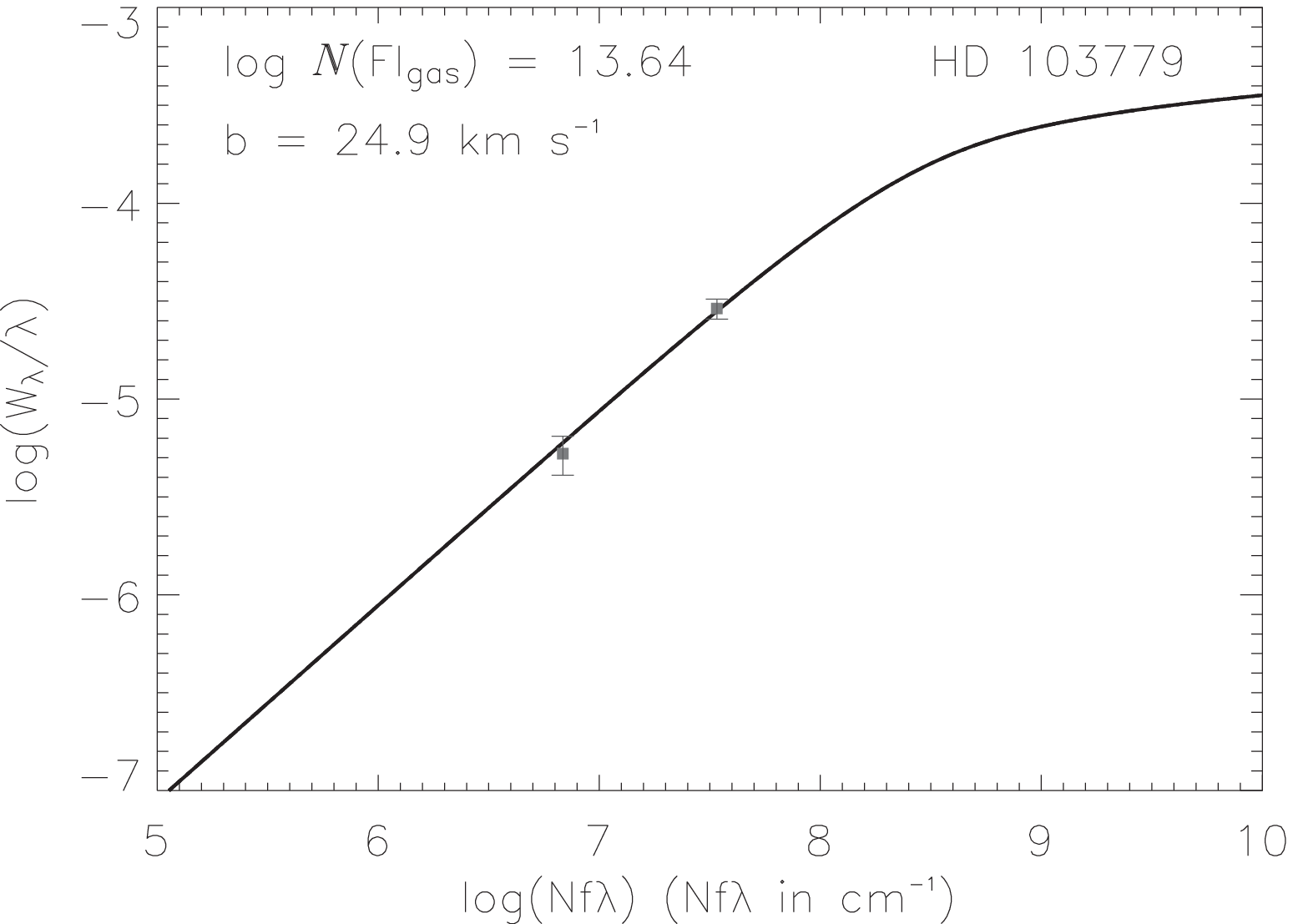} \quad \includegraphics{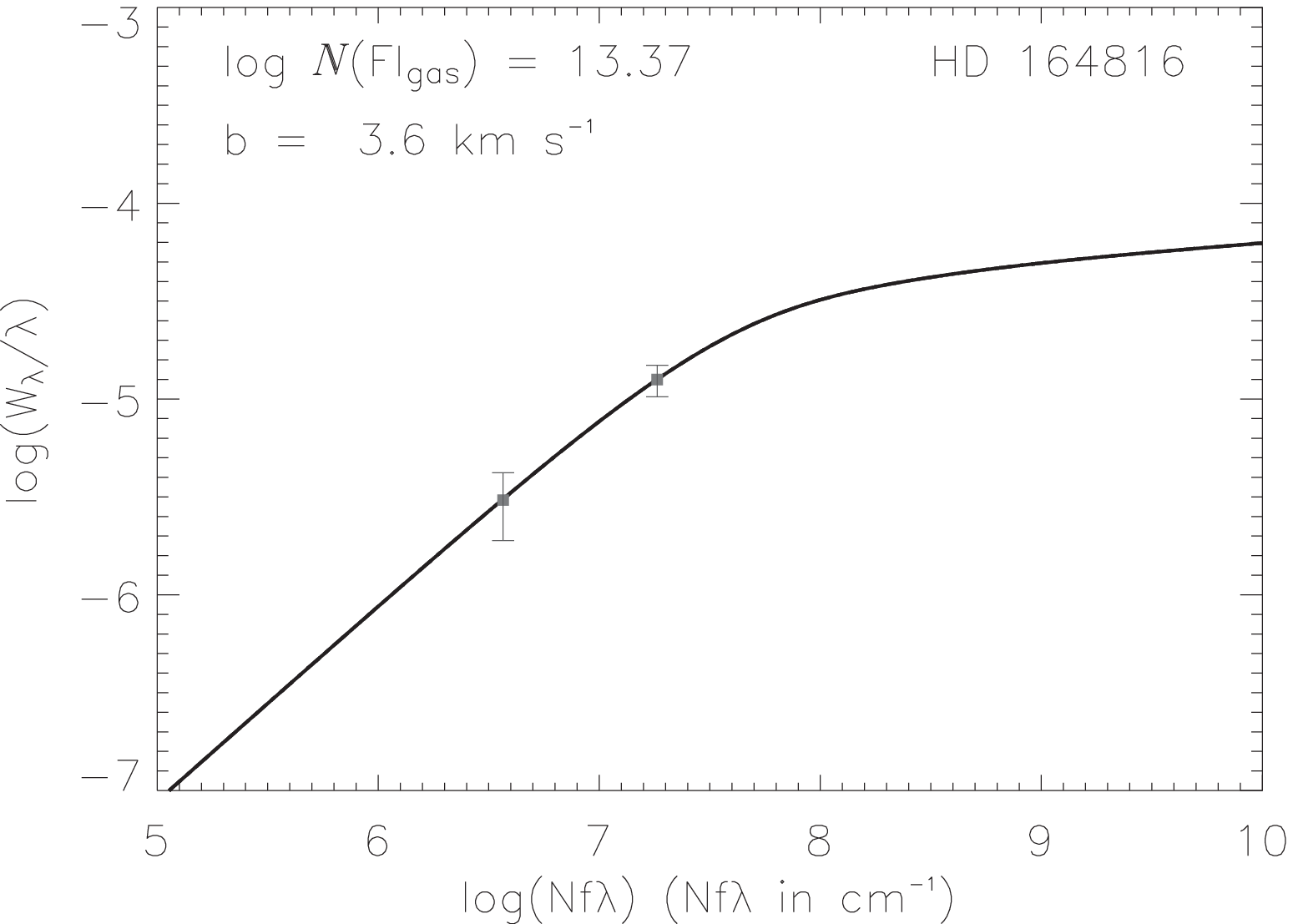}}

\vspace{.3 cm}

\scalebox{.5}[.5]{\includegraphics{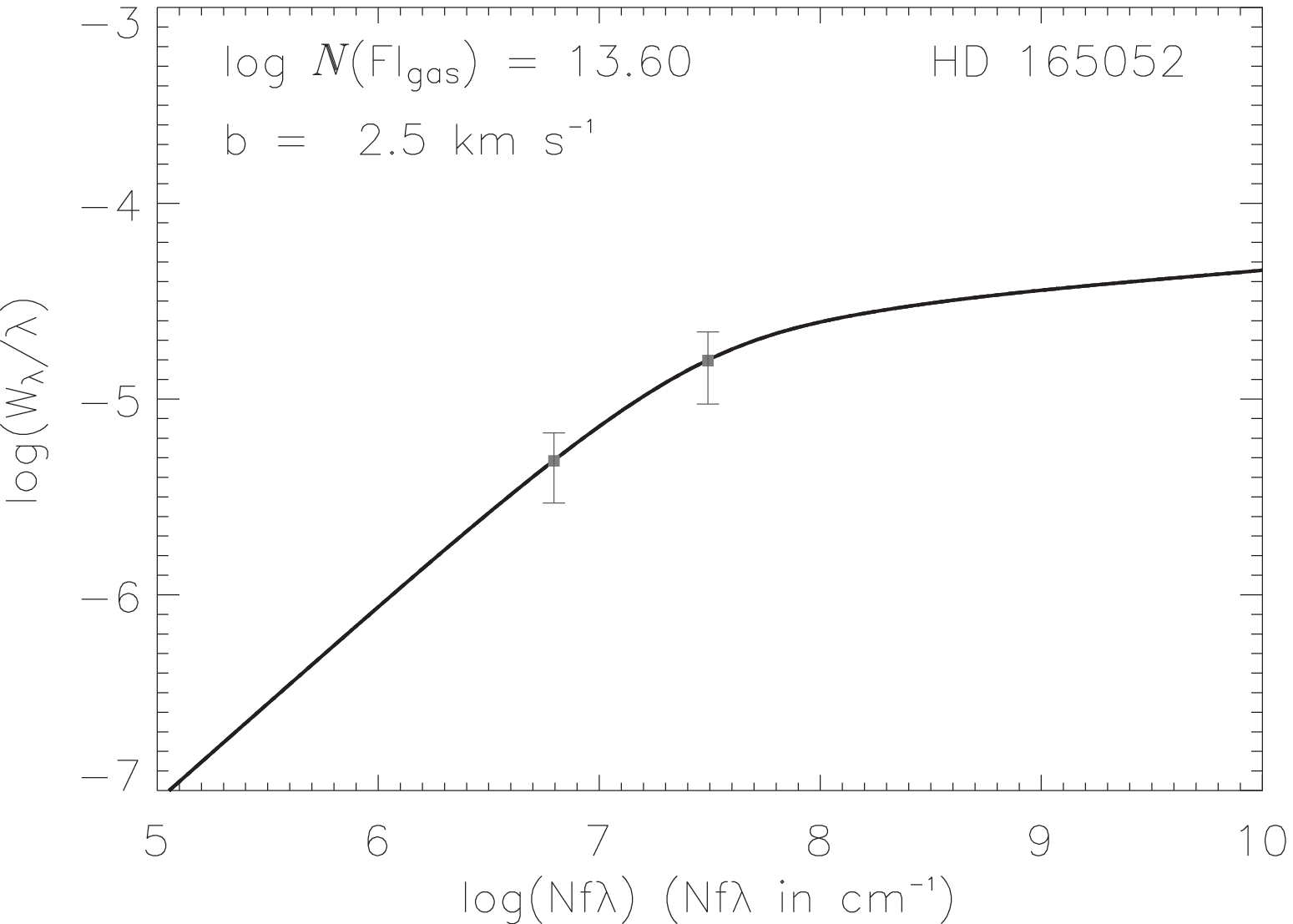} \quad \includegraphics{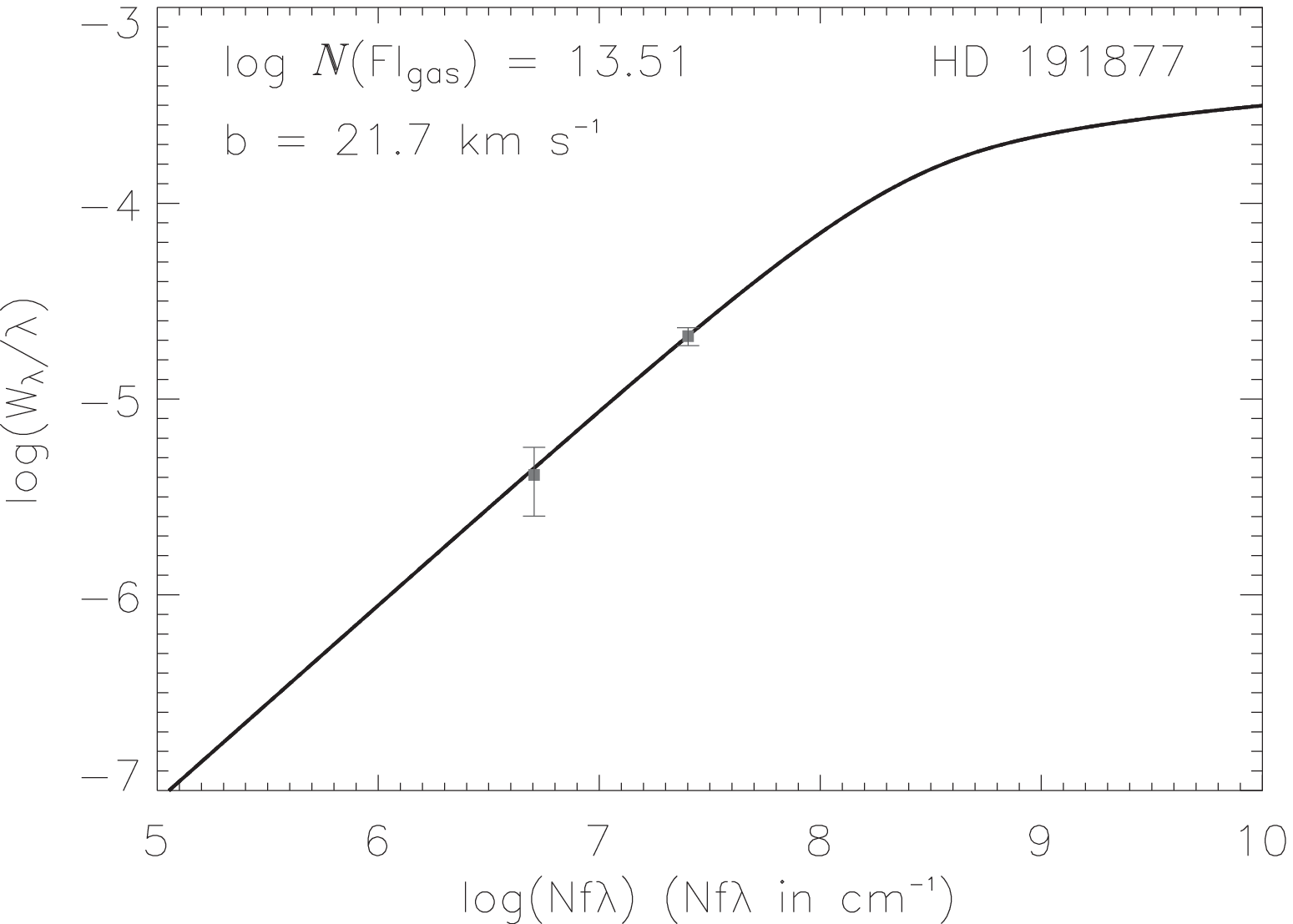}}

\caption{F I curves of growth for lines of sight where both F I lines were measured.  Measured equivalent widths and error bars are plotted on top of the appropriate curve for the derived column density and {\it b}-value.  In all cases we observe the ratio of F I equivalent widths to be consistent with little to no saturation.  This low level of saturation is also predicted by the N I curve of growth.\label{fig_Fcog}}
\end{center}
\end{figure}
\end{center}

For HD 103779, HD 164816, HD 165052, and HD 191877 both F I lines were measured with 
a single Gaussian to at least two-sigma significance.  So for each of these 
cases we derived a curve of growth based solely on the two F I lines 
(see Figure \ref{fig_Fcog}).  From these curves we calculated F I column densities 
and {\it b}-values and compared them to those acquired using the N I 
curve of growth (See Table \ref{tab_meathodcompare}).  The values derived from these 
two methods agree with one another.  
Since we only have two points on the curve, we cannot perfectly constrain its 
shape, so errors in derived column densities and {\it b}-values are obtained by 
letting each variable in turn change while the other is held at its derived optimal value.

HD 37367 had a strong absorption feature at the wavelength of the 951 \AA\ F I 
line (Figure \ref{fig_954_951_1}), and the wavelength shift of 
this line matched the shifts of nearby N I lines. However, this 
line of sight had very little flux in the region around the 954 \AA\ F I line, 
making the 954 \AA\ line immeasurable.  Additionally, HD 37367 has a 
low {\it vsin(i)} value ($\sim$ 20 km/s) making  stellar lines narrow and not easily 
distinguishable from interstellar features.  Thus, for this sightline we calculated 
the column density based on the 951 \AA\ feature, but because the spectra are so 
confusing, we cannot confidently say that the feature at 951 \AA\ 
is F I, especially without the detection of the 954 \AA\ feature.

One sightline through the Monoceros Loop supernova remnant (HD 47240) appeared 
to have a very strong absorption feature at the wavelength of the 951 \AA\ F I line.  
An analysis of the velocity structure for this line of sight revealed this feature 
to most likely be a high velocity ($\sim$ 70 km/s) component of the nearby 
H$_2$ J = 3 line.  

In our final analysis of F I column densities, we took the detections, 
or upper limits, on the F I 951 \AA\ line into account, which helped 
us to refine our errors and improve our confidence in the F I column 
densities.

\subsection{Discrepancies when Measuring the 954 \AA\ Line of F I}
In two lines of sight, HD 93205 and HD 103779, it is unclear whether the line 
at 954 \AA\ is F I or not.  In the spectrum of HD 93205, there is a 
large absorption feature at the wavelength of the 954 \AA\ 
F I line; however, we do not believe this feature to be F I but rather 
a high velocity ($\sim$ $-$90 km/s) component of the nearby H$_2$ J = 1 line.  
Thus, for HD 93205 we have not claimed a detection of F I but only 
assign a 2$\sigma$ upper limit.  For HD 103779 the absorption feature 
at 954 \AA\ is wider than in other lines of sight.  This is not surprising, 
though, as we find the N I {\it b}-value to be high (17.3 km/s) and high 
resolution observations show two or more components spanning about 
30 km/s in velocity space (Jenkins, \& Tripp, 2001).

As mentioned earlier, we expect F I to be similarly distributed 
in space as N I and we nearly always observe F I wavelength shifts to indeed match 
those of N I.  However, in three cases (HD 164816, HD 165052, and HD 315021) 
both F I lines' central wavelengths do not seem to correlate to those of N I.  
Interestingly, all three of these stars are located in 
NGC 6530, a young cluster and star forming region in the Lagoon Nebula (M 8).  
In two of these lines of sight (HD 164816 and HD 165052) we were able to 
measure both the 954 \AA\ and 951 \AA\ F I lines, making us confident that 
we are indeed measuring F I.  The appearance of the same shift in three different 
lines of sight confirms that this is not just coincidental to one line of 
sight but that the effect is real.  However, we are unable at this point 
to explain why F I would not have the same velocity 
structure as N I in these lines of sight.

\subsection{Fluorine Formation}
As mentioned in the introduction, one of the main objectives in our study of fluorine 
is to determine what mechanisms are primarily responsible for the element's 
formation.  Thus we attempted to explore this topic by investigating 
Wolf-Rayet stars and sight lines that pass through known supernova remnants.  
Unfortunately this did not shed much light on any of the possible formation mechanism 
because of poor data quality and unidentified, interfering absorption features.  
For those lines of sight that did have sufficient flux, we set upper limits 
on the equivalent width of the 954 \AA\ and 951 \AA\ F I lines and derived 
upper limits on the F I column densities. Results can be found in 
Tables \ref{tab_Feqwid}, \ref{tab_NColDen}, \ref{tab_FColDen}, and \ref{tab_WolfRayet}.


\section{Discussion}

We have detected F I absorption in 13 of our 26 targets, and provide upper 
limits for the remaining twelve, with HD 37367 ambiguous because of 
possible interference from narrow stellar features.  
Our range of reddening and total gas column density is 
restricted on both sides - at the lower limit by the minimum column density needed 
for the F I line to be detected, and on the upper side by the beginning of total 
obscuration of the F I line by molecular hydrogen absorption.  Thus we are 
unable to explore fluorine depletions over a very wide range of cloud physical 
conditions.  Any future studies will suffer the same limitations. 

Nevertheless we can reach some useful conclusions about the fluorine 
abundances and depletions in the diffuse sightlines that we were able 
to probe.  First, in general agreement with the results of Snow and 
York (1981) and of Federman et al. (2005) we find that fluorine is 
only slightly depleted, if at all,  relative to the solar value for 
all of the sightlines where we obtained detections of F I absorption.  
Our values for depletion (Table 5) show F/H ratios ranging from 
about 0.1 to -0.6 dex. 

Second, we do not see any sign of fluorine enhancement even in 
sightlines toward early O stars where the internal nucleosynthesis 
might have been expected to produce excess fluorine which was then 
expelled into the ISM through rapid stellar winds (i.e., the 
Wolf-Rayet mechanism mentioned in Section I).  It 
is impossible to obtain UV spectra toward AGB stars, so the best 
way to determine fluorine abundances in those environments is 
either through the J = 1-0 HF absorption at 243 $\mu$m, something that will become 
possible when the SOFIA aircraft is in operation (see discussion 
in Neufeld et al. 2005) or through the J = 2-1 HF absorption at 121 $\mu$m (Neufeld et al. 1997). 

One possible explanation, though an unlikely one, for the lack of observed 
fluorine enhancement in early O stars could be that extra fluorine 
expelled from these stars is depleted rapidly into molecules or dust grains.  
If this were happening, it might explain why we do not see 
significant levels of either fluorine depletion or enhancement.

The slight depletion of atomic fluorine requires two conditions: 
(1) not much of the fluorine in the diffuse ISM is in molecular 
form; and (2) there is little depletion of fluorine onto dust grains.  
These points are apparently in disagreement with the predictions of 
Neufeld et al. (2005), if they were applied to dense clouds. 

The hydride of fluorine, HF, is expected to compete with, or perhaps exceed, 
the abundance of atomic fluorine, depending on how abundant H$_2$ is. 
For molecular fractions approaching unity, HF should dominate over F I 
(Neufeld, Wolfire, and Schilke, 2005), but for H I regions, atomic fluorine should 
be dominant.  Our H$_2$ fractions are in range from 0.1 to 0.4, so we should 
expect more F I than HF, which we see (we know that, because F I is nearly 
solar, as alluded to above and as described below).  HF has been detected 
in emission toward the far-IR continuum source Sg2 B2 (at 121 $\mu$m; 
Neufeld et al. 1997), but the far-UV line near 951 \AA\ has not been seen, 
though there are hints that it may be contributing to an overlapping N I 
line (Sonnentrucker and Neufeld, private communication).   

Neufeld et al. (2005, 2006) stress that the abundance of HF is likely to 
peak just where the ambient photon field encounters gas containing 
substantial quantities of H$_2$, that is, at the edges of dense clouds 
or Photon Dominated Regions (PDRs).  Since our sightlines contain a 
mix of clouds and H I gas, we might expect just what we apparently 
are seeing:  F I is dominant over HF, but there still may be a bit of 
HF.  For now, the abundance of HF in our observed sightlines is unknown, 
but apparently less than F I. 

Another diatomic species, CF$^+$, could be strongly present, but 
always less so than HF (Neufeld, Wolfire, and Schilke, 2005). From 
mm-wave emission measurements of the Orion Bar, Neufeld et al. (2006) 
report $N_{CF^+} \approx\ 10^{10}$, 
whereas the far-UV data from Snow and York (1981), 
Federman et al. (2005), and this study show that atomic fluorine 
is three orders of magnitude larger.  The relative abundances of 
HF and CF$^+$ are consistent with the models of Neufeld et al. 
(2005, 2006), but apparently not the relative abundances of HF and F I. 

Regarding depletion onto grains, both Snow and York (1981) and 
Federman et al. (2005) have discussed this.  Snow and York, in 
the absence of any information on the condensation temperature of 
fluorine, would have attributed the modest depletion to the high 
ionization potential (I.P.) of fluorine, following Snow (1975), who 
found an inverse correlation between I.P. and depletion.  Recently, 
however, a condensation temperature was published by Lodders (2003), 
who gives a value for fluorine of about 730 K.  This should lead to 
substantial (on order of 0.8 to 1.0 dex) depletion if fluorine 
followed a correlation with condensation temperature (see review 
by Savage and Sembach 1996).  Neufeld et al. (2005) infer in their 
models a depletion of 1.7 dex. On the other hand, if the inverse 
correlation with I.P. holds, then with an I.P. of 17.422 eV, fluorine 
should be undepleted, more closely approximating what we see. 

It is not clear why fluorine is not bound onto the grains.  Fluorine 
is such a rare element that models of dust grains, extinction, and 
grain surface reactions usually ignore it.  But given the reactivity of 
atomic fluorine with molecular hydrogen, and given 
that interstellar grains are presumed to be coated with H$_2$, it would 
be seem that HF and other fluorine compounds should form on 
grains and that therefore a significant fraction of the extant fluorine 
should reside there. This, along with the gas-phase chemistry analyzed 
by Neufeld et al. (2005, 2006), combined with the detections of HF and 
CF$^+$ in PDRs, makes it likely that most of fluorine is in molecular 
form in dark clouds, as predicted.  But until the most likely molecular 
species, such as HF and CH$^+$, are detected in the same diffuse 
molecular lines of sight where F I is also observed, we remain unsure 
of the dominant state of fluorine in those regions. 

\acknowledgments
This research was supported by FUSE grant NNG05GA85G and COS grant NAS5-98043.  
We are grateful for several useful and stimulating comments made by the 
anonymous referee.  We would also like to thank Rachel Wiley for her 
assistance in editing this publication.


\clearpage

\clearpage

\clearpage

\begin{deluxetable}{llccccccr}
\tablecolumns{9}
\tablewidth{0pc}
\tabletypesize{\footnotesize}
\tablecaption{Stellar Data\label{tab_stardata}}
\tablehead{Star (HD)   &       Type    &       E(B-V)  &       V       &       Distance (kpc) & $N_{HI}$ & $N_{H_2}\tablenotemark{d}$ & $N_{H}$  &Molecular Frac}
\startdata

12323   &       B9	   &       0.29    &       8.92    &  2.29   &    21.18\tablenotemark{c} &	20.31  & 21.28 &   0.21     \\
37367   &       B2IV    &       0.40    &       5.99    &  0.36   &    21.28\tablenotemark{c} &	20.62  & 21.44 &   0.30     \\
39680   &   O6:pe       &       0.34    &       7.89    &  2.48   &    21.30\tablenotemark{a} &	19.53  & 21.31 &   0.03     \\
90087   &B2/B3III       &       0.28    &       7.80    &  2.72   &    21.15\tablenotemark{a} &	19.91  & 21.17 &   0.10     \\
93205   &        O3V    &       0.37    &       7.76    &  2.63   &    21.33\tablenotemark{a} &	19.79  & 21.34 &   0.06     \\
94493   &B0.5Iab/Ib     &       0.20    &       7.27    &  3.33   &    21.11\tablenotemark{a} &	20.12  & 21.15 &   0.17     \\
103779  & B0.5II	   &	   0.21	   &	   7.22	   &  4.07   &	  21.16\tablenotemark{a} &	19.83  & 21.20 &   0.09	    \\
116538  &       B2Ib/II &       0.17    &       7.88    &  1.11   &    21.04\tablenotemark{a} &	19.63  & 21.06 &   0.07     \\
116781  & B0:Iab:e      &       0.43    &       7.45    &  1.49   &    21.18\tablenotemark{b} &	20.06  & 21.24 &   0.13     \\
116852  &        O9III  &       0.22    &       8.49    &  4.80   &    20.96\tablenotemark{a} &	19.78  & 20.99 &   0.12     \\
152623  &        O8     &       0.44    &       6.68    &  1.91   &    21.28\tablenotemark{a} &	20.21  & 21.32 &   0.15     \\
164816  &        O+     &       0.31    &       7.09    &  1.59   &    21.18\tablenotemark{a} &	20.00  & 21.21 &   0.12     \\
164906  &        O+     &       0.42    &       7.45    &  1.59   &              	         &	20.22  &       &      -     \\
165052  & A5		   &   	   0.41	   &	   7.76    &  1.59   &    21.36\tablenotemark{a} &	20.20  & 21.42 &   0.12	    \\
177989\tablenotemark{e} & B0III  & 0.25 &       9.34    &  4.91   &    20.95\tablenotemark{a} &      20.12  & 21.06 &   0.23     \\
191877  &       B1Ibe   &       0.18    &       6.27    &  2.22   &    20.89\tablenotemark{a} &	20.02  & 20.95 &   0.21     \\
195965  &       B0V     &       0.25    &       6.98    &  0.79   &    20.90\tablenotemark{a} &	20.36  & 21.01 &   0.37     \\
202347  &       B1V     &       0.17    &       7.50    &         &    20.99\tablenotemark{b} &	19.98  & 21.07 &   0.16     \\
208440  &        B1V    &       0.34    &       7.91    &  0.62   &    21.23\tablenotemark{a} &	20.29  & 21.28 &   0.19     \\
209339  &       B0IV    &               &       6.69    &  0.85   &    21.16\tablenotemark{b} &	20.21  & 21.25 &   0.18     \\
315021  & B3V	   &	   0.31	   &	   8.59    &  1.59   &    21.28\tablenotemark{a} &	19.99  & 21.32 &   0.09	    \\

\vspace{-.3 cm}\\
\multicolumn{9}{l}{\bf{Wolf-Rayet Stars}}\\ 
\hline
\vspace{-.3 cm}\\
92809   &         WC6   &       0.22   &        9.08    &           &               	     	 &	20.23  &          &  -     \\
93162   &         WN+   &       0.62   &        8.11    &  2.60     &  21.55\tablenotemark{a} &	19.83  &  20.56   &  0.19  \\
191765  &         WN6   &       0.45   &        8.31    &           &  21.48\tablenotemark{a} &	20.27  &  21.51   &  0.06  \\
\vspace{-.3 cm}\\										       		 
\multicolumn{9}{l}{\bf{Stars Behind Supernova Remnants}} \\							       		 
\hline												       		 
\vspace{-.3 cm}\\										       		 
74711   &	B1III	   &	   0.33	  &	   7.11	   &	       &		         &	20.30  &	  &  -	   \\
74920   &        O+     &       0.34   &        7.54    &  1.50     &  21.15\tablenotemark{a} &	20.26  &  21.20   & 0.21   \\

\enddata
\tablenotetext{a}{H I data taken from \citet{DiplasSavage}.}
\tablenotetext{b}{H I data taken from Jensen et al. (in preparation).}
\tablenotetext{c}{H I data taken from \citet{Cartledge}.}
\tablenotetext{d}{$H_2$ data were taken from Shull et al. (in preparation).}
\tablenotetext{e}{HD 177989 lies behind the Scutum Supershell.}

\end{deluxetable}

\begin{deluxetable}{lrrr}
\tablecolumns{4}
\tablewidth{0pc}
\tabletypesize{\small}
\tablecaption{New H$_2$ Column Densities \label{tab_H2colden}}
\tablehead{Star (HD)	& $N_{J = 0}$  & $N_{J = 1}$  & $N_{J = 2}$}
\startdata

37367	&  20.36 $\pm$ 0.04 & 20.29 $\pm$ 0.07  & 18.64 $\pm$ 0.25  \\
74711	&  20.08 $\pm$ 0.05 & 19.97 $\pm$ 0.07  & 18.32 $\pm$ 0.15  \\
92809	&  19.88 $\pm$ 0.06 & 19.94 $\pm$ 0.05  & 18.87 $\pm$ 0.10  \\
93162	&  19.40 $\pm$ 0.03 & 19.68 $\pm$ 0.04  & 18.54 $\pm$ 0.11  \\
94493	&  19.85 $\pm$ 0.06 & 19.78 $\pm$ 0.02  & 18.55 $\pm$ 0.17  \\
116538	&  19.24 $\pm$ 0.05 & 19.44 $\pm$ 0.02  & 18.30 $\pm$ 0.16  \\
164906	&  19.96 $\pm$ 0.02 & 19.88 $\pm$ 0.02  & 17.95 $\pm$ 0.25  \\
191765	&  19.91 $\pm$ 0.11 & 19.95 $\pm$ 0.12  & 18.76 $\pm$ 0.13  \\
202347	&  19.66 $\pm$ 0.05 & 19.72 $\pm$ 0.06  & 18.26 $\pm$ 0.16  \\
208440	&  20.04 $\pm$ 0.05 & 19.96 $\pm$ 0.02  & 18.39 $\pm$ 0.06  \\ 
209339	&  19.83 $\pm$ 0.03 & 19.96 $\pm$ 0.04  & 17.79 $\pm$ 0.16  \\ 
315021	&  19.57 $\pm$ 0.03 & 19.78 $\pm$ 0.02  & 18.08 $\pm$ 0.16  \\ 

\enddata
\end{deluxetable}

\begin{deluxetable}{lrr}
\tablecolumns{3}
\tablewidth{0pc}
\tabletypesize{\small}
\tablecaption{F I $W_{\lambda}$\label{tab_Feqwid}}
\tablehead{Star (HD)	& $F_{954}$ $W_{\lambda}$ (m\AA )& $F_{951}$ $W_{\lambda}$ (m\AA )}
\startdata
12323  &  $\leq$ 59\tablenotemark{a}  &	$\leq$ 20\tablenotemark{a}  \\
37367	& $\leq$ 78\tablenotemark{a} 	&  20 $\pm$	7	\\
39680  & 20 $\pm$	4 	      & $\leq$ 11\tablenotemark{a}   \\
90087  & 15 $\pm$	3  		&   $\leq$ 6\tablenotemark{a}       \\
93205  &  $\leq$ 30\tablenotemark{a}	      & $\leq$ 8\tablenotemark{a}   \\
94493  & 20 $\pm$	8   & $\leq$ 13\tablenotemark{a}    \\
103779 & 21 $\pm$	3	& 5.0 $\pm$	1.1	\\
116538 & 8 $\pm$	4  & $\leq$ 6\tablenotemark{a}    \\
116781 &  $\leq$ 17\tablenotemark{a}  & $\leq$ 10\tablenotemark{a}    \\
116852 &  $\leq$ 10\tablenotemark{a}  & $\leq$ 7\tablenotemark{a}    \\
152623 &  $\leq$ 27\tablenotemark{a}  & $\leq$ 14\tablenotemark{a}    \\
164816 &  12 $\pm$	2  		&  2.9 $\pm$	1.1      \\
164906 &  $\leq$ 55\tablenotemark{a}  & $\leq$ 17\tablenotemark{a}    \\
165052 &  15 $\pm$	6  		&  4.6 $\pm$	1.8      \\
177989 &  18 $\pm$	7 	& $\leq$ 9\tablenotemark{a}	\\
191877 & 20 $\pm$	2 	      & 3.9 $\pm$ 1.5   \\
195965 & $\leq$ 23\tablenotemark{a} 	& $\leq$ 3\tablenotemark{a}   \\
202347 & 7 $\pm$	3 	      & $\leq$ 6\tablenotemark{a}   \\
208440 & 19 $\pm$	7 	      & $\leq$ 15\tablenotemark{a}   \\
209339 & 16 $\pm$	6 	      & $\leq$ 13\tablenotemark{a}   \\
315021 & 18 $\pm$	5 	      & $\leq$ 6\tablenotemark{a}   \\
\vspace{-.3 cm}\\
\multicolumn{3}{c}{\bf{Wolf-Rayet Stars}}\\ 
\hline
\vspace{-.3 cm}\\
92809	& $\leq$ 37\tablenotemark{a} 	& $\leq$ 36\tablenotemark{a}	\\
93162	& $\leq$ 30\tablenotemark{a} 	& $\leq$ 27\tablenotemark{a}	\\
191765	& $\leq$ 44\tablenotemark{a} 	& $\leq$ 22\tablenotemark{a}	\\
\vspace{-.3 cm}\\
\multicolumn{3}{c}{\bf{Supernova Remnants}} \\
\hline
\vspace{-.3 cm}\\
74711	& $\leq$ 30\tablenotemark{a}	&  $\leq$ 24\tablenotemark{a}	\\
74920  &  $\leq$ 58\tablenotemark{a}  & $\leq$ 11\tablenotemark{a}    \\

\enddata
\tablenotetext{a}{Two sigma upper limit}
\end{deluxetable}

\begin{deluxetable}{lrr}
\tablecolumns{3}
\tablewidth{0pc}
\tabletypesize{\small}
\tablecaption{Interstellar N I Column Densities and {\it b}-values\label{tab_NColDen}}
\tablehead{Star (HD)	&$N_{NI}$ &{\it b}-value (km s$^{-1}$)}
\startdata
12323   & 17.03 &  5.8    \\
37367   & 17.60 & 5.9	\\
39680   & 16.90 &  7.1      \\
90087   & 16.68 &  9.9    \\
93205   & 16.87 &  17.0     \\
94493   & 17.01 &  9.7    \\
103779  & 16.70 &  17.3     \\
116538  & 16.74 &  13.1   \\
116781  & 17.01 &  12.2   \\
116852  & 16.91 &  10.8   \\
152623  & 17.05 &  21.8   \\
164816  & 16.67 &  9.4    \\
164906  & 16.81 &  10.8   \\
165052  & 16.85 & 15.1	     \\
177989  & 16.71 & 9.4     \\
191877  & 16.76 &  7.8    \\
195965  & 16.71 &  5.5    \\
202347  & 16.52 &  8.4    \\
208440  & 17.01 &  8.9    \\
209339  & 17.16 &  5.8    \\
315021  & 16.51 &  14.7   \\

\vspace{-.3 cm}\\
\multicolumn{3}{c}{\bf{Wolf-Rayet Stars}}\\ 
\hline
\vspace{-.3 cm}\\
92809	& 17.75	& 10.7	\\
93162	& 17.79	& 16.3	\\
191765	& 17.24	& 10.4	\\
\vspace{-.3 cm}\\
\multicolumn{3}{c}{\bf{Supernova Remnants}} \\
\hline
\vspace{-.3 cm}\\
74711	& 17.34 & 8.6 \\
74920   & 16.93 &  11.1   \\

\enddata
\end{deluxetable}

\begin{deluxetable}{lll}

\tablecolumns{3}
\tablewidth{0pc}
\tabletypesize{\footnotesize}

\tablecaption{Interstellar F I Column Densities and Depletions\label{tab_FColDen}}
\tablehead{Star	(HD)&$N_F$ & Depletion\tablenotemark{b}}
\startdata

12323   & $\leq$ 14.57\tablenotemark{a}  	& $\leq$ +0.73  \tablenotemark{a}                \\
37367   & $14.32^{\ 0.19}_{-0.24}$\tablenotemark{c}	& +0.32\tablenotemark{c}  	\\
39680   & $13.58^{\ 0.09}_{-0.12}$      	& -0.29			\\
90087   & $13.39^{\ 0.10}_{-0.13}$  		& -0.34    	    \\
93205   & $\leq$ 13.71\tablenotemark{a}    	& $\leq$ -0.19  \tablenotemark{a}	\\
94493   & $13.55^{\ 0.17}_{-0.25}$	  	& -0.16     	    \\
103779  & $13.53^{\ 0.06}_{-0.05}$	   	& -0.23 \\
116538  & $13.10^{\ 0.18}_{-0.30}$		& -0.52   	    \\
116781  & $\leq$ 13.46\tablenotemark{a}  	& $\leq$ -0.34 \tablenotemark{a}     	    \\
116852  & $\leq$ 13.21\tablenotemark{a}      & $\leq$  -0.34  \tablenotemark{a}	    \\
152623  & $\leq$ 13.65\tablenotemark{a}  	& $\leq$  -0.23  \tablenotemark{a}   	    \\
164816  & $13.29^{\ 0.08}_{-0.08}$   	& -0.48    	    \\
164906  & $\leq$ 14.11\tablenotemark{a}  	& -				 	    \\
165052  & $13.39^{\ 0.15}_{-0.23}$	   	& -0.59    \\
177989  & $13.49^{\ 0.17}_{-0.23}$		& -0.13	\\
191877  & $13.57^{\ 0.04}_{-0.06}$      	&  0.08			\\
195965  & $\leq$ 13.69\tablenotemark{a}  	& $\leq$  0.12  \tablenotemark{a}		\\
202347  & $13.05^{\ 0.16}_{-0.25}$     	& -0.58			\\
208440  & $13.52^{\ 0.17}_{-0.21}$     	& -0.32			\\
209339  & $13.47^{\ 0.18}_{-0.23}$      	& -0.34			\\
315021  & $13.47^{\ 0.12}_{-0.14}$ 	  	& -0.41   \\

\vspace{-.3 cm}\\
\multicolumn{3}{c}{\bf{Wolf-Rayet Stars}}\\ 
\hline
\vspace{-.3 cm}\\
92809	&  $\leq$ 13.86\tablenotemark{a}	& -	\\
93162	&  $\leq$ 13.71\tablenotemark{a}	&$\leq$ 0.59\tablenotemark{a}	\\
191765	&  $\leq$ 13.97\tablenotemark{a}	&$\leq$ -0.10\tablenotemark{a}	\\
\vspace{-.3 cm}\\
\multicolumn{3}{c}{\bf{Supernova Remnants}} \\
\hline
\vspace{-.3 cm}\\
74711	& $\leq$ 13.77\tablenotemark{a}	&	-	\\
74920   & $\leq$ 14.14\tablenotemark{a}  & $\leq$ +0.38  \tablenotemark{a}     	    \\

\enddata
\tablenotetext{a}{Two sigma upper limit}
\tablenotetext{b}{Depletions are based on the solar value F/H = $3.63 \times 10^{-8}$.}
\tablenotetext{c}{Values based on the 951 \AA\ F I line.  This F I detection is uncertain due to potential interference from stellar features.}
\end{deluxetable}

\begin{deluxetable}{lrr}
\tablecolumns{3}
\tablewidth{0pc}
\tablecaption{Comparison With Previous Work\label{tab_Federmancompare}}
\tablehead{Star (HD)	&$N_F$ $_{(This\ Study)}$	&$N_F$ $_{(Federman)}$}
\startdata
208440	&$13.52^{\ 0.17}_{-0.21}$	&$13.52^{\ 0.16}_{-0.26}$ \\
209339	&$13.47^{\ 0.18}_{-0.23}$	&$13.46^{\ 0.17}_{-0.28}$ \\
\enddata
\end{deluxetable}

\begin{deluxetable}{lcccr}
\tablecolumns{5}
\tablewidth{0pc}
\tablecaption{Comparison of Methods used to Derive $N_F$  \label{tab_meathodcompare}}
\tablehead{Star (HD)& \multicolumn{2}{c}{$N_F$}& \multicolumn{2}{c}{{\it b}-value (km s$^{-1}$)} \\
	& N I & F I & N I & F I}
\startdata
103779	&$13.67^{\ 0.05}_{-0.05}$       &$13.64^{\ 0.04}_{-0.04}$       &17.3       &$24^{\ 0}_{-13}$	\\
164816	&$13.29^{\ 0.08}_{-0.08}$       &$13.37^{\ 0.10}_{-0.09}$       &9.6     &$3^{\ 16}_{-3}$	\\
165052	&$13.39^{\ 0.15}_{-0.23}$       &$13.60^{\ 0.22}_{-0.19}$       &15.1       &$2^{\ 22}_{-1}$	\\
191877	&$13.57^{\ 0.04}_{-0.06}$       &$13.51^{\ 0.04}_{-0.05}$       &7.8       &$21^{\ 3}_{-14}$	\\

\enddata
\end{deluxetable}

\begin{deluxetable}{lr}
\tablecolumns{2}
\tablewidth{0pc}
\tabletypesize{\small}
\tablecaption{Observed lines of sight to Wolf-Rayet stars or through supernova remnants \label{tab_WolfRayet}}
\tablehead{Star (HD)	& Problem with line of sight}

\startdata
\multicolumn{2}{c}{\bf{Wolf-Rayet Stars}}\\ 
\hline
\vspace{-.3 cm}\\
	
32402	&	Interference from other absorption features	 \\
33133	&	Noise and insufficient flux in 954 \AA\ region	 \\
37026	&	Interference from other absorption features	 \\
37680	&	Interference from other absorption features	 \\
96548	&	Insufficient flux in 954 \AA\ region		 \\
104994	&	Noise and insufficient flux in 954 \AA\ region	 \\
151932	&	Insufficient flux in 954 \AA\ region		 \\
164270	&	Insufficient flux in 954 \AA\ region		 \\
187282	&	Interference from other absorption features	 \\
192163	&	Insufficient flux in 954 \AA\ region		 \\
192641	&	Noise and insufficient flux in 954 \AA\ region	 \\
269582	&	Noise and insufficient flux in 954 \AA\ region	 \\

\vspace{-.3 cm}\\
\multicolumn{2}{c}{\bf{Supernova Remnants}} \\
\hline
\vspace{-.3 cm}\\
5980    &	Interference from other absorption features	   \\
36665   &	Insufficient flux in 954 \AA\ region		   \\
37318   &	Insufficient flux in 954 \AA\ region		   \\
47240   &	High velocity components interfere with 951 \AA\ region \\
72088   &	Noise and insufficient flux in 954 \AA\ region	   \\
72350   &	Noise and insufficient flux in 954 \AA\ region	   \\

\enddata

\end{deluxetable}

\end{document}